\documentclass[conference]{IEEEtran}
\pdfoutput=1
\usepackage{amssymb,amsmath}
\usepackage{subfigure}
\usepackage[lflt]{floatflt}
\usepackage{cite}
\usepackage[dvips]{color}
\usepackage{epsfig,graphicx}
\usepackage{euler}
\usepackage{calc,ifthen}
\usepackage{rotating}
\usepackage{bbm, dsfont, mathbbol, mathrsfs, eufrak}
\DeclareMathAlphabet{\mathpzc}{OT1}{pzc}{m}{it}
\usepackage{wrapfig}
\usepackage{algorithmic}
\usepackage{algorithm}
\usepackage{verbatim}
\usepackage{multicol}

\newcommand{\T}{\mathcal{T}}

\newcommand{\rr}{\mathcal{O}}
\newcommand{\Xx}{\mathcal{X}}

\newcommand{\bu}{\pmb{u}}
\newcommand{\bv}{\pmb{v}}
\newcommand{\bh}{\pmb{h}}

\newcommand{\zero}{\mathbf{0}}

\let\hat\widehat

{\end{list}}

\newcommand{\entrylabel}[1]{\mbox{\textnormal{#1}}\hfil}%
\newenvironment{entry}%
   {\begin{list}{}%
       {\renewcommand{\makelabel}{\entrylabel}%
        \settowidth{\labelwidth}{\textnormal{Return values}} %
        \setlength{\leftmargin}{\labelwidth+\labelsep}%
       }%
   }%
   {\end{list}}

\newcommand{\Mentrylabel}[1] %
   {\raisebox{0pt}[1ex][0pt]{\makebox[\labelwidth][l]%
       {\parbox[t]{\labelwidth}{\hspace{0pt}\textnormal{#1}}}}}
   {\renewcommand{\entrylabel}{\Mentrylabel}\begin{entry}}%
   {\end{entry}}

\definecolor{subtler}{rgb}{1,0,0.1}  
\definecolor{subtleb}{rgb}{0.1,0,0.9}  

\newcommand{\be}{\begin{equation}}
\newcommand{\ee}{\end{equation}}
\newcommand{\ba}{\begin{eqnarray}}
\newcommand{\ea}{\end{eqnarray}}

\def\lossvec{\mathbf{b}}

\def\pathloss{\mathbf{p}}

\newcommand{\x}{\pmb{x}}
\newcommand{\z}{\pmb{z}}

\newcommand{\y}{\pmb{y}}

\newcommand{\norm}[2]{\Vert #1 \Vert_{#2}}
\newcommand{\gup}{\text{UpSparse}^+}

\def\xmin{x_{\min}}
\def\ymin{y_{\min}}
\def\yumin{y^u_{\min}}
\def\yumax{y^u_{\max}}
\def\ylmin{y^\ell_{\min}}
\def\ylmax{y^\ell_{\max}}
\def\argymin{m_{\min}}
\def\argxmin{m_{\min}}

\def\ylj{y^\ell_j}
\def\yL{y^L}
\def\yU{y^U}

\def\yuj{y^u_j}

\def\zui{z^u_i}
\def\zuj{z^u_j}
\def\zli{z^L_i}
\def\zlj{z^L_j}
\def\zmi{z^\ell_i}

\def\llog{\cal L}
\newcommand{\up}{\text{UpSparse}}

\def\3up{65mm}
 
\def\oneup{35mm} 
\def\Oneup{80mm}

\def\fourup{38mm} 

\def\nontrivialtree                  {\includegraphics[width=\oneup]{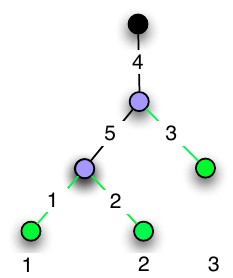}} 
\def\exampleKonetreea        {\includegraphics[width=\fourup]{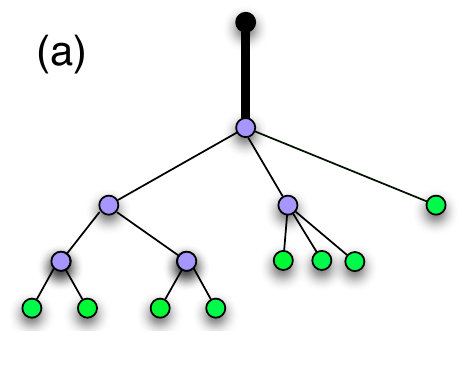}}  
\def\exampleKonetreeb        {\includegraphics[width=\fourup]{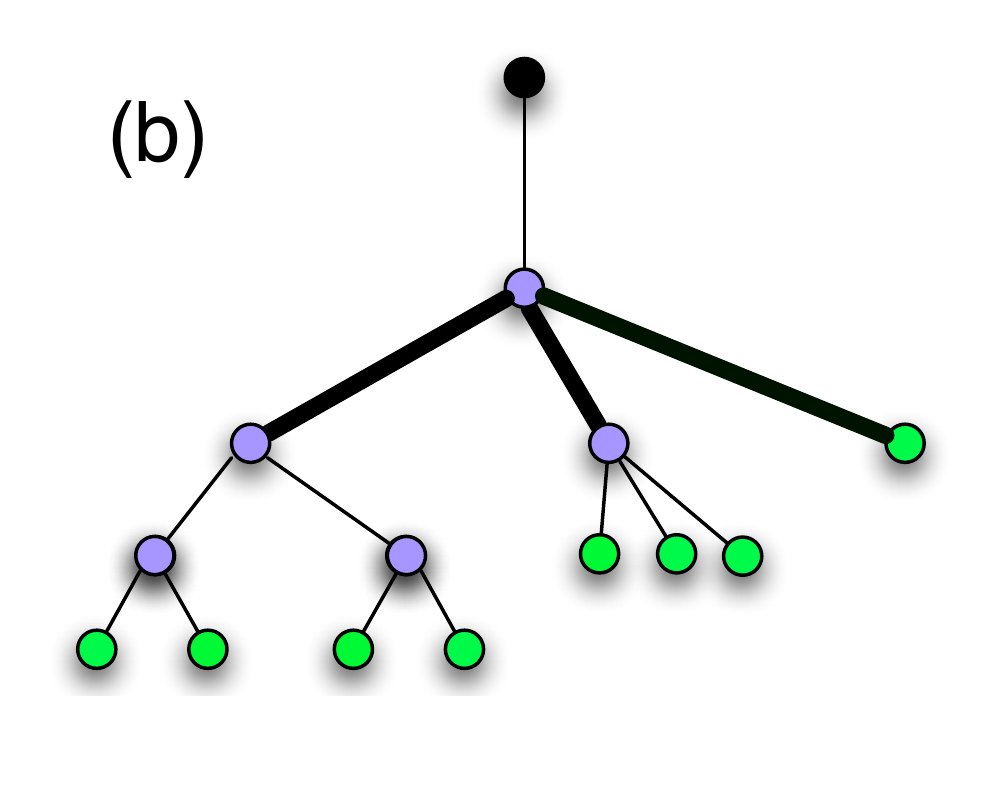}}  
\def\exampleKonetreec        {\includegraphics[width=\fourup]{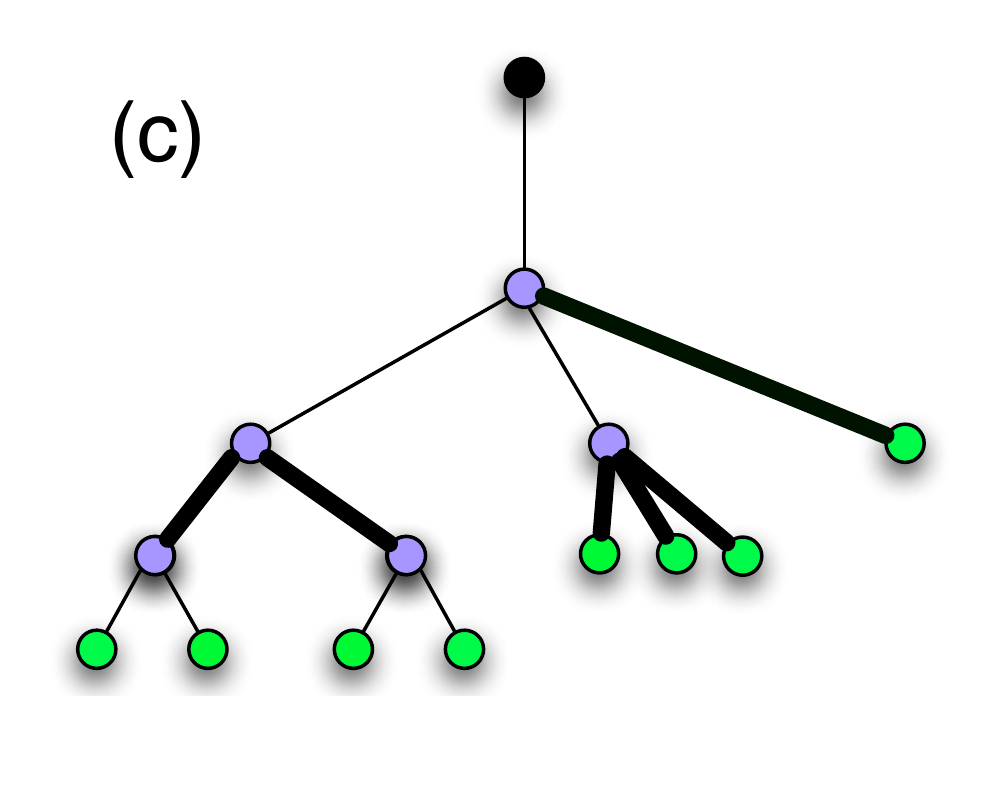}}  
\def\exampleKonetreed        {\includegraphics[width=\fourup]{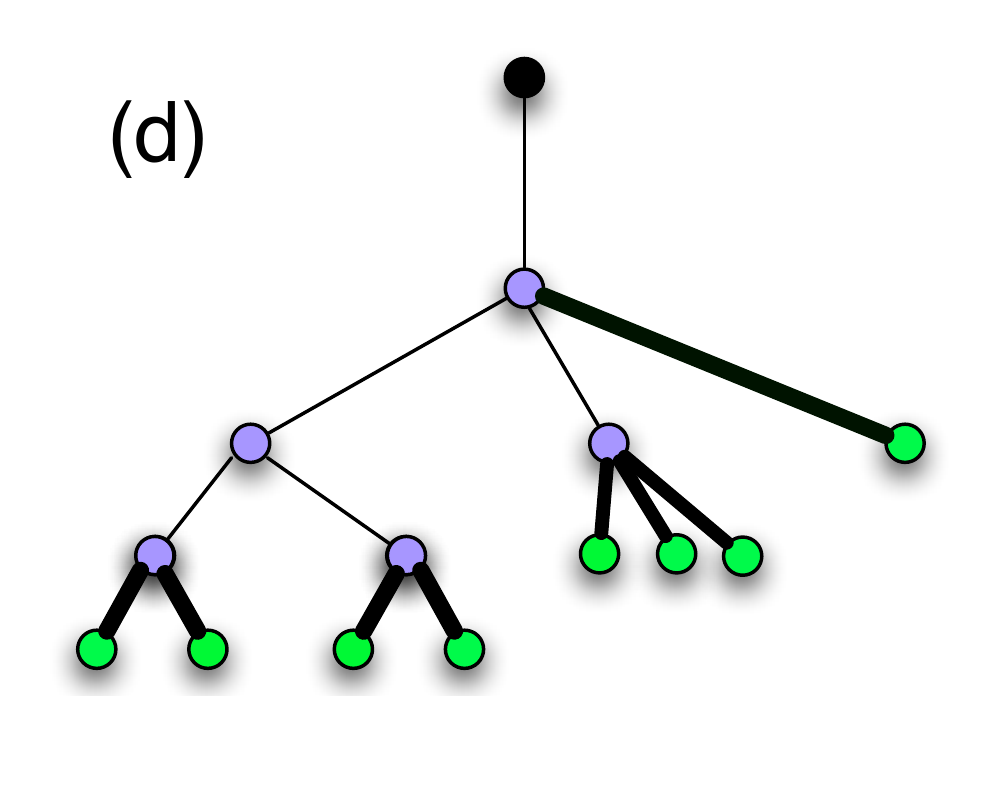}}  

\def\localtree                             {\includegraphics[width=\oneup]{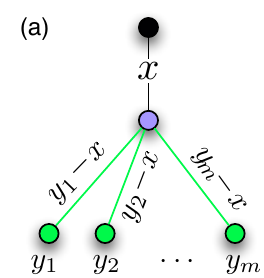}} 
\def\coupledlocaltree                 {\includegraphics[width=\Oneup]{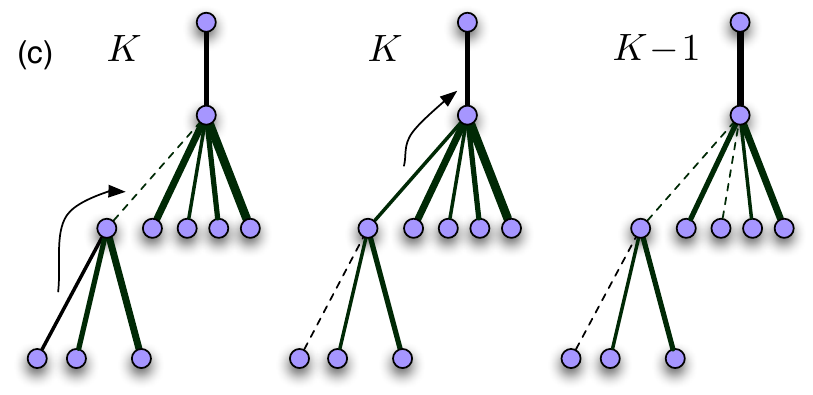}}

\begin{document}

\title{Sparsity without the Complexity: Loss Localisation using Tree Measurements\vspace{-0.3cm}}
\author{
\IEEEauthorblockN{Vijay Arya}
\IEEEauthorblockA{IBM Research -- India\\
Bangalore, India\\
Email: vijay.arya@in.ibm.com}
\and
\IEEEauthorblockN{Darryl Veitch}
\IEEEauthorblockA{
Department of Electrical and Electronic Engineering\\ The University of Melbourne, Australia\\
Email: dveitch@unimelb.edu.au}
}

\maketitle

\begin{abstract}
We study network loss tomography based on observing average loss rates over a set of paths forming a tree -- a severely underdetermined linear problem for the unknown link loss probabilities. 
We examine in detail the role of sparsity as a regularising principle, pointing out that the problem is technically distinct from others in the compressed sensing literature.
While  sparsity has been applied in the context of tomography, key questions regarding uniqueness and \mbox{recovery} remain unanswered. 
Our work exploits the tree structure of path measurements to derive sufficient conditions for sparse solutions to be unique and the condition that $\ell_1$ minimization recovers the true \mbox{underlying} solution. 
We present a fast single-pass linear algorithm for $\ell_1$ minimization and prove that a minimum $\ell_1$ solution is both unique and sparsest for tree topologies. 
By considering the placement of lossy links within trees, we show that sparse solutions remain unique more often than is commonly supposed.
We prove similar results for a noisy version of the problem.

\end{abstract}


\section{Introduction}

Network operators and end applications alike would  like to localize abnormally lossy links or \textit{loss hotspots}, but how can this be achieved when internal access to the network is limited?
Consider a set of nodes instrumented as active probing sources or receivers, generating flows of probes over a set of paths in the network to measure loss.
The intersections of these paths result in a set of relations for mutual consistency of the measured path loss probabilities in terms of the constituent link loss probabilities that one would like to recover.  This is a network tomography problem,    defined over the measurement sub-network traversed by the probes, which can be expressed as a linear system.  This system is in general severely under-determined: instead of a unique solution for the link  loss rates, an entire family of solutions is consistent with the observed path measurements.
%

One way to select a particular solution from the family, that is one regularising principle, is \textit{sparsity}: preferring the solution with the smallest number of lossy links.  
Sparsity is in keeping with Occam's razor which seeks the simplest explanation to a given set of observations, and is a natural fit to the assumption that hotspots are rare.  It also sits well with an operational need to provide a short list of potential hotspots worthy of closer attention.
This paper examines in detail the role of sparsity in network loss tomography in the important special case of a tree-like measurement topology or \textit{measurement matrix $A$}.

Trees have been considered in a few prior works treating sparsity in loss tomography (see section~\ref{sec:related}), but mostly in an implicit sense, as a default special case of general networks. 
In this paper we show that the structure of trees can be exploited to allow an essentially complete picture to be obtained including  conditions for uniqueness of sparse solutions, the relationship of sparse to $\ell_1$ solutions, and the condition that minimum $\ell_1$ recovers the true underlying solution.  

Tree measurement topologies arise in several practical contexts -- for instance unicast path measurements between a web server and its clients form a tree. 
Thus information on locations of hotspots could be used to direct clients to replica servers or have the hotspots resolved in cooperation with the concerned ISP~\cite{venkat}.  
Moreover, since any general measurement infrastructure can be configured for use as a tree, our results are relevant in practice. 
Furthermore,  loss rates inferred from multiple intersecting trees over the same infrastructure can be used to 
quickly obtain important partial information about 
general measurement topologies/matrices. 
Exploiting tree solutions and insights to gain purchase on the more general problem is a new direction and the subject of ongoing work.  An early result in this direction is Lemma~1 in section~\ref{ssec:suffnet}. 

There are three main differences between our problem setting and that of compressed sensing (CS), which has in recent years exploited sparsity for signal recovery (see Table~\ref{tab:CStreecompare}). 
First, network tomography deals with positive quantities such as link loss probabilities and delays, whereas CS  generally treats real valued signals. 
Second, in CS one inquires after the nature of the measurement matrix $A$, and its size (number of observations $m$) needed to recover a solution of given sparsity $K$ uniquely. 
In network tomography both the nature and size of $A$ are highly constrained by the availability of measurement nodes,  and the lack of control over the packet routing between them. 
In the case of a tree, adding a new `measurement' is non-trivial as it implies installing an active probe receiver in a new location. 
Finally, the measurement matrices studied in CS are designed (or assumed) to satisfy strong technical conditions such as the restricted isometry property (RIP \cite{decoding, Cand_Romb_Tao_06a}).  
In contrast, for a tree the measurement matrix $A$ is given rather than designed, and has a specific dependence structure. 
%
\begin{table*}
\hfil  \begin{tabular}{|c || c  | c|}
   \hline
   ~ & \textbf{Classical CS} (sparsity $K$) &  \textbf{Loss Tree Results} (sparsity $K$)\\ \hline\hline    
  Signal $x$    & $\mathbb{R}^n$ & link loss vector ${\mathbb{R}^+}^{\!n}$ \\\hline
   $m$ measurements & $m \ll n$, variable  & $m = n(1 - 1/c) + o(1)$ (for a $c$-child tree), \textbf{fixed} \\\hline  
  Matrix $A$    & $m\times n$ entries each $\sim \mathcal{N}(0,1/m)$     &  $m\times n$ binary matrix representing an arbitrary tree \\\hline\hline 
 Uniqueness   & every $2K$ columns of $A$ independent  &  Every branch node has at least 2 lossless incident links ($K \le m$)\\\hline
 Efficient Recovery       & $m = O(K \log (n/k))$, RIP conditions on $A$ &  Every branch node has at least 1 lossless child link ($K \le m$)\\\hline
         \hline
 \end{tabular}
 \caption{A comparison of problem statements and results for a typical CS problem compared to loss tomography over a tree.  The tree problem is more specific and detailed, tied to the structure of the tree rather than technical assumptions.
} 
 \label{tab:CStreecompare}
  \vspace{-9mm}
\end{table*}

Studying sparse solutions over trees is quite different from `traditional CS' approaches, however, the key questions of interest are the same: {\bf hardness, uniqueness, and recovery}. 
%
%
Compared to prior CS work, our task is nontrivial in that we do not benefit from properties such as RIP and must develop fresh techniques, but easier in that the structure of trees is simple and powerful.  As we see below, the net result is that the tree context is more tractable, enabling detailed and complete solutions with desirable properties. Our main contributions include:\\
\textbf{Hardness (Complexity of computing a sparse solution)}: 
Recovering the sparsest or minimal $\ell_0$ (pseudo) norm solution of under-determined systems is in general NP-hard \cite{natarajan}. We show that for a tree the sparsest solution(s) may be characterised precisely, and either written down explicitly or found with a fast linear time algorithm.

\noindent\textbf{Uniqueness (Conditions for sparsest solution to be unique):} 
Provided the number of lossy links at every internal/branch node with node degree $g$ is at most $g\!-\!2$, the sparsest solution is unique.  
This result takes into account the locations of loss within the tree, and shows that uniqueness may hold for much higher values of sparsity $K$ than the worst case analysis typically used in CS 
would suggest.   When solutions of a given sparsity are not unique, the alternative solutions can be precisely localised and characterised.

\noindent\textbf{Recovery (Conditions that minimum $\ell_1$ solution is the true underlying solution): }  
For the general problem, the minimal $\ell_1$ norm solution does not always have the minimal $\ell_0$ norm and need not even be unique. 
For the tree problem, we show that minimal $\ell_1$ solution always has the minimal $\ell_0$ norm and is unique. 
Provided every internal/branch node has at least one lossless child link, the minimal $\ell_1$ recovers the true underlying solution. 
We define the `UpSparse' algorithm, a fast single-pass linear-time algorithm which outputs the minimal  
$\ell_1$ solution. For the general problem, the minimal $\ell_1$ solution is recovered through a linear program (cubic complexity).

Since in practice only a finite number of probes can be sent, the measured loss probabilities can only be known approximately.
We formulate and study a `noisy' version of the  problem that addresses this key practical concern.
As before, we exploit the nature of the tree, rather than reusing regularisation approaches from other contexts (see section~\ref{sec:comp}). 
We characterise the minimal norm solutions and present fast algorithms to recover them. 
We observe that, unlike the noiseless case, the minimal $\ell_1$ solution is no longer unique and need not always have the minimal $\ell_0$ norm. 

We begin by discussing related work in section~\ref{sec:related}. 
Section~\ref{sec:model} presents the general solution for the hotspot localisation problem in trees. 
Sections~\ref{sec:regularize} and \ref{sec:noise} characterise the sparse and minimal $\ell_1$ solutions for noiseless and noisy problems.   
Having established what the sparse solutions are and how to find them, section~\ref{sec:comp} compares our algorithms with CS optimization techniques.
Section~\ref{sec:exp} presents experimental results where we explore the relationship between the sparse and true solutions.

\section{Related Work}
\label{sec:related}
A number of different problems~\cite{venkat, binary, chen, krig, netquest, netscope, minc,  rabbat}, all related to inferring link parameters using either unicast or multicast path measurements, have been studied under the purview of Network Tomography. 
Whereas multicast measurements utilise observations at the per-probe level,
the unicast 
tomography problem works with average observations of paths and reduces to an under-determined linear system in terms of unknown link parameters. 
However the most common approach to solving an underdetermined system, namely choosing the minimal $\ell_2$ norm solution, may not be suitable when the link quantity is concentrated at particular locations. For example, for loss inference, it tends to spread the loss  over all links in the network.

More recent work borrows techniques of recovering sparse solutions to underdetermined linear systems from compressed sensing (CS)~\cite{donoho06, decoding, Cand_Romb_Tao_06a}. 
In CS, a fixed but randomly generated (generally Gaussian) matrix $A$ is used to `measure' an unknown signal $\x$ as $\y_{m\times 1} = A_{m \times n}\,\x_{n \times 1}$; $m \ll n$ so that $x$ is underdetermined. 
CS results show that a minimally sparse $\x$ can be recovered with high probability (i.e.~for most $A$) using $\ell_1$ minimization \cite{bp}. 
As outlined above, network loss tomography is quite different: $\y$ represents the path observations, $\x$ the unknown link parameters, and $A$, which determines which paths traverse each link, cannot be chosen freely and has unknown properties in general.

Despite these differences, $\ell_1$ minimization has been used as a black box to recover sparse solutions in tomography. In~\cite{netquest}, Bayesian experimental design is used to determine the set of paths to measure in a network and a variant of $\ell_1$ minimization is used to infer link parameters. In~\cite{netscope}, variance in path measurements across multiple measurement intervals is used to identify a prior, and an $\ell_1$ minimization formulation from~\cite{netquest} is used to find a sparse solution close to it. 
In~\cite{venkat}, path measurements between a server and its clients are used to recover link loss rates by using sampling, Bayesian inference, and a variant of $\ell_1$ minimization. 
None of these works provide insight into the nature of the sparse or $\ell_1$ solutions, how they interact, or their uniqueness, the central focii of our work. 
In~\cite{binary}, locations of 'bad' network links is inferred from path measurements forming a tree. 
For this, each path is classified as 'good' ($0$) or 'bad' ($1$) and the smallest set of bad tree links consistent with the binary path observations is recovered. 
%
In section~\ref{sec:exp}, we see that this two-step approach fails to recover the true locations of hotspots more often than our approach that directly recovers a minimally sparse link loss solution. In addition, we recover both the locations and loss rates of lossy links. 
%
Given a measurement matrix, \cite{globecom} uses expander graphs~\cite{indyk} to determine conditions for recovering unique sparse  solutions in networks. 
However, for trees expander graphs do not bring any additional insight.

The work most closely related to our own is \cite{infocom}, which answers some of the 
key questions for CS over graphs.
The key difference is that we work with trees instead of general networks.
The simpler tree topology enables far greater insight into the sparse and  $\ell_1$ solutions, and allows explicit solutions and fast algorithms to be defined. 
In~\cite{infocom} the authors determine the number of random measurements over underlying network paths needed to uniquely recover sparse link solutions. Random measurements however are difficult to justify in the tomography context. Conversely, for a given measurement matrix they provide upper bounds on the number of lossy links  consistent with uniqueness of the sparsest solution.
These bounds are quite restrictive for trees. For example for any ternary tree, irrespective of its size, the largest allowed number of lossy links is $2$, and for a binary tree the price of a uniqueness guarantee is that only a single link may be lossy. 
In section~\ref{sec:exp}, we see that for a ternary tree with $25$ links, even when $4$ links are lossy,  the sparsest solution is still unique for $95\%$ of feasible link loss vectors, and the proportion grows with tree size. In section~\ref{ssec:suffnet}, we also show how the recovery of a $K$-sparse vector relates to the degree of the measurement graph.

\section{The Noiseless Hotspot Problem and Solution}
\label{sec:model}

In this section we describe how we model the loss process over a tree,
and how to formulate the resulting problem as a linear system. 
We then solve the system formally, and make some preliminary observations.

 \vspace{-1mm}
\subsection{Model Definition}
\label{ssec:defn}

\noindent\textbf{Tree Model} \quad
Let $\T = (V,L)$ denote the logical tree consisting of a set of nodes $V$ and links $L$. 
Let $\rr \in V$ denote the root node, $R \subset V$ be the set of leaf nodes, 
and $I=V\setminus\{\rr\cup R\}$ the set of internal nodes. 
A link is an ordered pair $(j, k) \in \{V \times V\}$ representing a logical link (one or more physical links) from node $j$ to node $k$. 
For each node $k$ except the root there is a unique node $j = f(k)$, the father of $k$, such that $(j, k) \in L$. 
The set of children of a node $j$ is denoted by $c(j)$, thus $c(j)=\{k \in V: (j,k) \in L\}$. 
All nodes have at least two children, except the root (just one) and the leaves (none). 
The depth of a node is the number of links in the (unique) path of ancestors leading to the root. 
By level $l$ of a tree we mean the set of nodes of depth $l$, with the root being of depth zero.
We denote the \textit{height} of the tree (the depth of the deepest leaves) by $H$.
The \textit{top link} is the unique link adjacent to the root node.

For convenience, we refer to link $(f(k), k)$ simply as link $k$, and similarly, we also use $I$ to refer to the set of \textit{internal links} corresponding to the internal nodes,  
$R$ to refer to the leaf links as well as nodes, and so on.
Let $n$ denote the number of links in the tree, $m = |R|$ the number of leaves, and $d=n-m$ the number of internal links. From each leaf there is a unique path to the root, 
so $m$ is also the number of paths. Clearly $n\ge m+1$.
It is convenient to label nodes/links as follows:  First, the leaf nodes are labelled by $k = 1, 2\ldots m$ from left to right.  
Then beginning with the child of the root the counting continues in a \textit{preorder} traversal of the internal nodes of the tree (recursively: node, left subtree right subtree). 
With this convention, the labels of leaf nodes can double as convenient path labels. In other words, path $j$ terminates at leaf node $j$, with paths labelled as $j = 1, 2\ldots m$ from left to right.  Examples are given in the figures.

The topology of the tree is captured by the $m\times n$ 
measurement matrix $A$, where entry $A_{jk}=1$ if link $k$ forms part of the path $j$, zero otherwise. 
Row $j$ of the matrix gives the links in path $j$, and column $k$ give the paths which cross link $k$.  

\noindent\textbf{Modelling Link Loss} \quad
The marginal probability of loss on link $k$ is given by $b_k\in[0,1]$, and we denote the $(n\times 1)$ vector of loss probabilities over all links by $\lossvec$. We assume stationarity so that $\lossvec$ is constant.
As in all prior work, spatial independence is assumed, 
i.e., 
all link loss processes are mutually independent.   It follows that the path loss probability is easily expressed via the product of the link passage probabilities:  $p_j = 1- \prod_{k:A_{jk}=1} (1- b_k)$ where the product is over the links on path $j$. 
We assume that we have access, 
through measurements based on a large number of probes, to the exact path loss probability vector, 
$\pathloss=[p_1, p_2, \ldots, p_m]^T$.

 \vspace{-1mm}
\subsection{System Solution}
\label{ssec:soln}

Define the \textit{addloss function}  as ${\cal L}( b) = -\log(1- b),~ b\in[0,1)$. 
We write $x_k = {\cal L}( b_k)$ and $y_j={\cal L}(p_j)$.   
Since $\cal L$ is a monotonically increasing function, mapping $[0,1)$ to $[0,\infty)$,  the link loss vector $\lossvec$ is replaced by the equivalent link \textit{addloss vector} $\x = [x_1,x_2,\ldots, x_n]^T$, and similarly $\pathloss$ is replaced by $\y=[y_1,y_2,\ldots, y_m]^T$.  
The relation $p_j = 1- \prod_{k:A_{jk}=1} (1- b_k)$ is now $y_j = \sum _{k:A_{jk}=1} x_k$,  and the relationship between path and link loss takes the linear form
\be
   \label{eqn:system}
   \y =  A \x  ,  \quad x_k\ge0, y_j\ge0.
\ee   
The term `addloss' is justified by the additive nature of link addloss, together with the fact that values of $x_k$ and $y_j$ can still be interpreted directly as loss for many purposes. 
In particular $0$ addloss implies zero loss.
Since we use addloss exclusively in this paper, we will use `loss' as a shorthand for addloss.

\begin{figure}[ht]
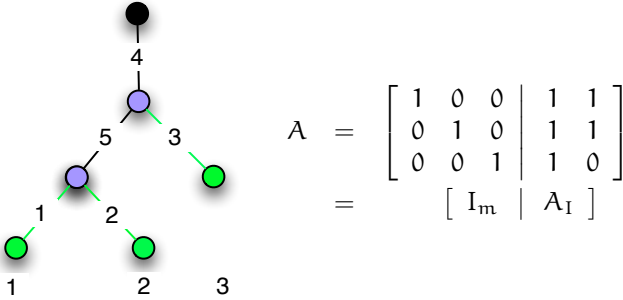

\vspace{0mm}
\centering
\begin{minipage}{0.42\linewidth}
\centering
 \hspace{-8mm}
   \nontrivialtree
\end{minipage} 
\begin{minipage}{0.46\linewidth}
\centering
\begin{eqnarray*}
\nonumber
A &=& \left[\begin{array}{ccc}
	1 &   0   &  0  \\
	0 &   1   &  0  \\
	0 &   0   &  1   
	\end{array}\right|
	\left.\begin{array}{cc}
          1  &   1   \\
          1  &   1   \\
	 1  &   0   
	\end{array}\right]\\
&=&  \hspace{8.2mm}
      \left[\begin{array}{c | l}
	I_m ~&~  A_I
	\end{array}\right] 	
\label{eqn:Anontrivial}
\end{eqnarray*}
\end{minipage}
\caption{A tree with $3$ leaves (and paths) and $2$ internal links, with its measurement matrix $A$.
Receiver links correspond to the identity matrix $I_m$.}
\vspace{-0.2cm}
\label{fig:nontrivialtree}
\end{figure}

\def\gap{\hspace{4mm}}
\begin{figure*}[ht]
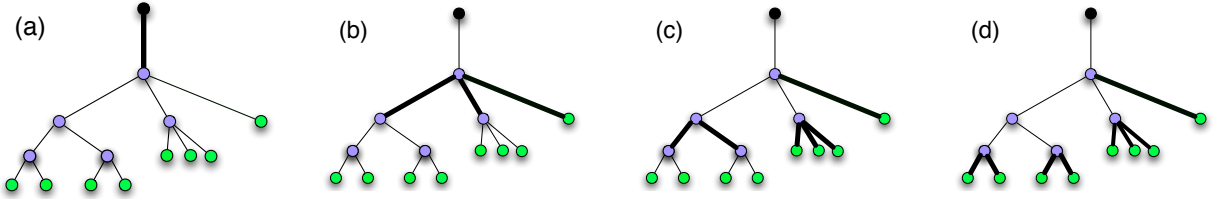

\centering
\exampleKonetreea \gap \exampleKonetreeb \gap \exampleKonetreec \gap\exampleKonetreed
\vspace{-5mm}
 \caption{An example of ambiguity in the locations of lossy links (bold), each of loss $x$. 
 In each case the receiver vector is $\y = [x, \dots, x]^T$, but the number of lossy links varies - 1, 3, 6, and 8.
 }
 \label{fig:exampleKonetree}
\vspace{-0.7cm} 
\end{figure*}

Consider the tree in figure~\ref{fig:nontrivialtree} together with its measurement matrix $A$, where the vertical divider separates the $m$ columns corresponding to receiver links $\x_R = [x_1, x_2, \ldots, x_m ]^T$ on the left, from those of the internal links $\x_I = [x_{m+1}, \ldots, x_n ]^T$ on the right.  
The $(m\times m)$ identity matrix $I_m$ in the left appears because each leaf link belongs to just one path (and because of our link and path naming conventions). This is true in general for any tree, and we may partition any measurement matrix into $I_m$ and the $(m\times d)$ matrix $A_I$ that shows how internal links contribute to paths. We can now rewrite (\ref{eqn:system}) as
\begin{align}
\nonumber
   \y &= A\x  
             = {\left[\begin{array}{cl}  I_m &  A_I	\end{array}\right] \atop }   \hspace{-2mm}
			\left[\begin{array}{c}
			   \x_R \\ 
			   \x_I
			\end{array}\right]         
             = \x_R  + A_I \x_I \\
\Rightarrow
	   \x(\x_I;\y) &= \left[\begin{array}{c}
				   \x_R \\
				   \x_I
			        \end{array}\right]
			        = \left[\begin{array}{c}
				   \y - A_I \x_I \\
				   \x_I
			        \end{array}\right] 
\label{eqn:gensoln}			                
\end{align}

It is clear that $A$ has full rank $m$, as its column rank is clearly at least $m$ due to the embedded identity matrix, and row and column rank are equal.
The general solution will therefore have $d=n-m$ free parameters.
Since there are also $d$ internal links, it is convenient to select $\x_I$ to span this space, in terms of
which a formal solution can be immediately written as above where the path observation vector $\y$ appears as a parameter.

The choice of  $\x_I$ as the independent variables has the advantage of making $\y$ appear in a simple way in the solution. 
For e.g., setting $\x=0$, it is clear that $\y=\zero$ is the corresponding observation.
Setting $\x_I=0$, we see that 
$\y=\x_R$, which also reveals that any set of (non-negative) observations is possible in general.  
We call  $\x_I=0, \x_R=\y$, the \textit{receiver solution}.

\section{Regularizing the Solution using Sparsity}
\label{sec:regularize}

Equation~(\ref{eqn:gensoln}) is a $d$-dimensional family of loss solutions all equally consistent with a single observed $\y$. 
From the point of view of an observer whose end goal is to identify a unique set of candidate loss hotspots, this represents a significant and problematic \textit{ambiguity}. 
Our main regularising principle is that of \textit{sparseness}, i.e. minimizing the number $K$ of lossy links which are consistent with any observed $\y$.  
A smaller number is preferred because it is more likely under a priori assumption that loss is rare, 
and focusses attention on a smaller number of candidate problem links, which has practical advantages. 
If in fact solutions are sparse, i.e. given a bound on the sparsity $K$, we wish to determine the conditions under 
which a sparsest 
solution is the unique solution consistent with observed $\y$.

Consider figure~\ref{fig:exampleKonetree} which gives four (of several) possible solutions consistent the observations $\y = [x,\ldots,x]^T$, $x>0$. 
Figure~\ref{fig:exampleKonetree}(a) shows the solution $\x_I=[x,0,\ldots,0], \x_R=0$ where only the top link is lossy while figure~\ref{fig:exampleKonetree}(d) shows the receiver solution $\x_I=0, \x_R=\y$. If sparsity $K \le 1$, then \ref{fig:exampleKonetree}(a) is the unique sparsest solution consistent with $\y$.

Finding the sparsest solution is equivalent to minimizing the $\ell_0$ (pseudo) norm of $\x$: $K = ||\x||_0= \sum_{i=1}^n |x_i|^0 =  \sum_{i:x_i>0} 1$.  
Results from CS have shown that minimizing with respect to $\ell_1$ norm often identifies solutions with minimal $\ell_0$ norm but at a lower computational cost.
We therefore also explore $\ell_1$ below, both in its own right, and as a secondary principle which can be used to further reduce ambiguity.

Whilst a priori information on the likely locations of lossy links within the topology may be available in some contexts, this is not always so.  In this paper we treat the case where there is no such information, corresponding informally to a uniform prior over all links.

\subsection{Local Regularisation}
\label{ssec:local}

\begin{figure}[h!]
\centering
   \localtree  \ 
   \includegraphics[width=45mm]{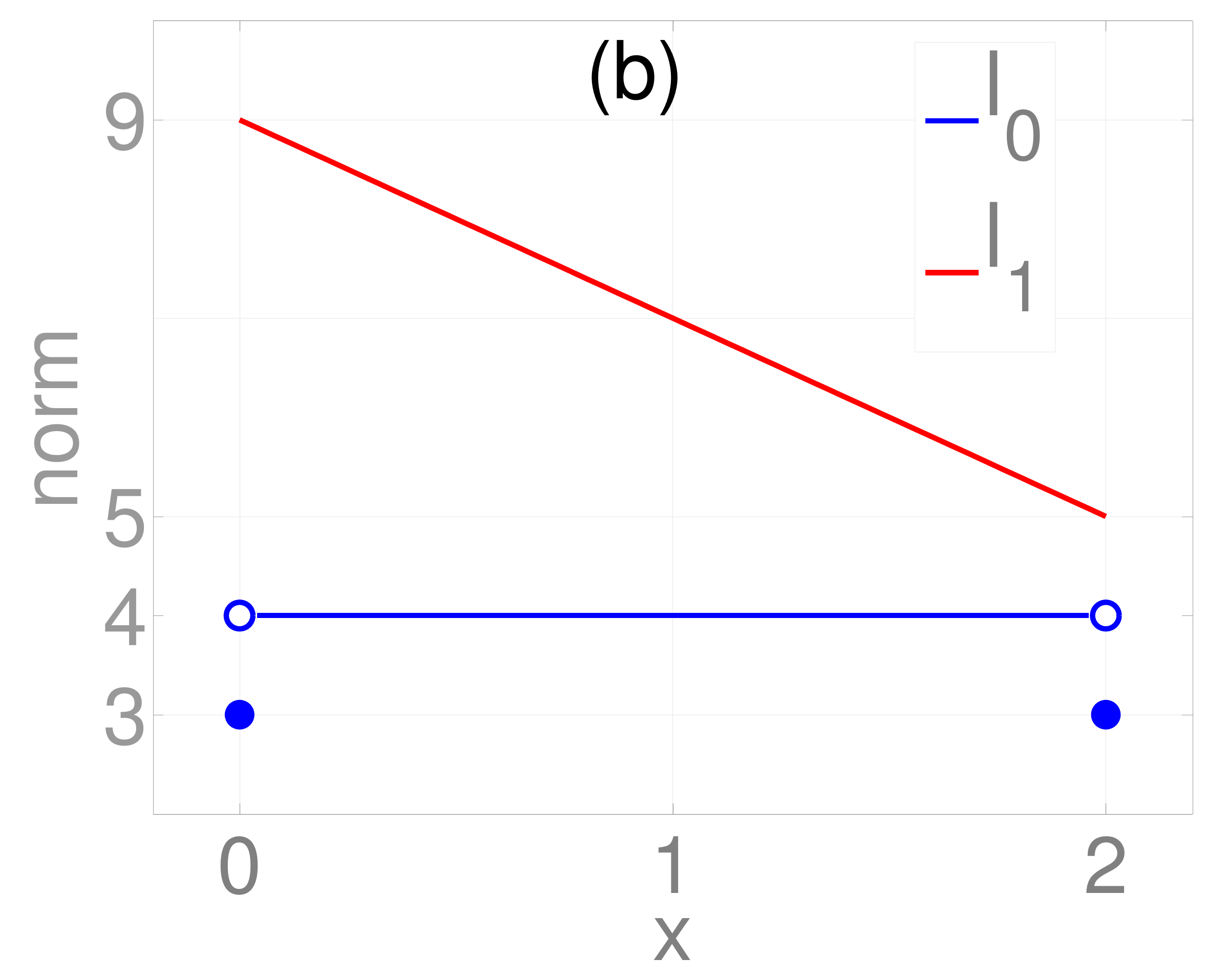}  
   ~\coupledlocaltree
\vspace{-2mm}   
 \caption{(a) Ambiguity at a local complex, which could be centered on any internal node, illustrated by a simple tree.
 Links and paths are annotated with their loss values, obeying the general solution parameterized by $x$. Loss can be moved up or down subject to $x\in[0,\ymin]$.
 (b) $\ell_0$ and $\ell_1$ norms of solutions shown for a 3-receiver complex with observations $\y = [2, 3, 4]$. Loss can be moved up or down subject to $x \in [0, \ymin=2]$. Both downstate and upstate solutions at $x=0$ and $x=2$ have the equal sparsity $K = m = 3$, but upstate achieves minimal $\ell_1$. 
  (c) An example of the coupling of two local complexes resulting in lower global ambiguity (link thickness proportion to loss, dashed links lossless). Left: initial configuration: lower complex in downstate, upper in upstate.  Middle: system remains $K$-sparse when the lower complex changes state, but upper complex is no longer minimally sparse. Right:  moving the upper complex into the upstate yields lower sparsity.}
 \label{fig:localtree}
\vspace{-2mm}   
\end{figure}

It is useful to understand ambiguity in a \textit{local complex}.  This is a `building block' consisting of an internal/branch node and its adjacent links. 
We will explore it using the two level tree of figure~\ref{fig:localtree}(a), where link and path labels have been dropped in favor of their loss values.

The general solution corresponding to the observed $\y$ is 
\be
 \x' = [y_1-x, y_2-x, \ldots, y_m-x,  x]^T,
\ee
 parameterised by $x\in[0,\ymin]$, where $\ymin=\min_j y_j$ denotes the smallest path loss. We examine sparsity in the local complex as a function of the parameter $x\in[0,\ymin]$.

$\ymin=0$:  the family collapses to a unique solution, $x=0$. 

$\ymin>0$:  $x$ is not uniquely determined by $\y$ and we speak of an \textit{ambiguous complex}.
At $x=0$ the internal link is lossless and the child link losses are maximized. We call this the \textit{downstate}, and it has sparsity $K=m$.
As $x$ increases over the range $x\in(0,\ymin)$ loss is `pulled up' equally and in parallel from each child link to the internal link. In these \textit{mixed states} all links are lossy and sparsity is  $K=m+1$, the largest possible. This upward transfer of loss ceases at the \textit{upstate} when $x=\ymin$, where suddenly the loss on all $\argymin$ links sharing the minimum value $\ymin$ becomes zero, and $K$ drops to $K=m-\argymin +1$.  

To summarise, requiring sparsity excludes the mixed states, singling out the downstate ($x=0$) and upstate ($x=\ymin$). 
If $\ymin$ is not unique ($\argymin>1$) then the upstate solution is the unique sparsest solution. 
For example when moving from figure~\ref{fig:exampleKonetree}(b) to figure~\ref{fig:exampleKonetree}(a), $K$ drops locally around the top internal node from 3 to 1 as the complex moves from the downstate to the upstate.
If instead $\argymin=1$, then both the upstate and downstate have sparsity $K=m$, so ambiguity, though greatly reduced, remains. 

\def\gapp{\hspace{15mm}}

If instead of sparsity we consider $\ell_1$  the conclusions are very similar but ambiguity vanishes.
The $\ell_1$ norm of the local complex illustrated in figure~\ref{fig:localtree}(a) is given by
\be
   ||\x'||_1= \sum_{k=1}^n |x_k| = \sum_{j=1}^m y_j - x(m-1) .
   \label{eqn:L1local}
\ee
It is clearly minimized by setting $x$ as large as possible, namely $x=\ymin$, i.e.~the upstate.
Hence for the $\ell_1$ norm upstate is always both optimal and unique, whereas for $\ell_0$ it is always optimal but not always unique. A simple example with $\y = [2,3,4]$ is shown in Figure~\ref{fig:localtree}(b).

\subsection{Global Regularisation}
\label{ssec:global}

The local ambiguity above can occur centered on any internal node in the tree.
A natural question is whether global effects may resolve the local ambiguities,  or alternatively result in new forms of  inherently global ambiguity. 
Consider the scenario shown in figure~\ref{fig:localtree}(c) where the original sparsity is $K$.
The lower complex is initially in a downstate (left) with $\argymin=1$ and the upper complex is in the upstate.
After the lower complex moves to the upstate, the sparsity remains at $K$ (middle), however the upper complex is no longer minimally sparse.  
Let $\xmin$ denote the minimum loss of the child links in the upper complex \textbf{after} the move, and $\argxmin'$ its associated multiplicity. Since $\xmin>0$ and $\argxmin'=2>1$ sparsity can be reduced to $K+1-\argxmin'=K-1$ by moving the upper complex to the upstate.

There are two important observations to make from the above example. 
First, choosing the upstate always achieves minimal local sparsity.
Second, choosing the upstate locally may also enable sparser states to be found in complexes higher in the tree.   
Together these motivate the following 
algorithm 
which defines a global solution
based on the systematic exploitation of local sparsity with a preference for the upstate solution in case of local non-uniqueness.  
Recall that local complexes are centered on internal nodes, so exist at levels $l=1$, $2\ldots$ $H-1$. 

\vspace{1mm}
\noindent\textbf{UpSparse Algorithm}\\
\textit{Begin with an arbitrary feasible solution $\x$.
For all ambiguous local complexes at the deepest level select the up sparse state.
Move up to the next level and repeat. Terminate at level 1. }

\setlength{\arrayrulewidth}{1.3pt}
\begin{figure}[h]
\begin{tabular}{l}
\hline
\begin{minipage}{0.45\textwidth}
\vspace{2mm}
\textbf{Function} {$UpSparse(\x)$}   \quad  \% arbitrary feasible soln $\x$
\begin{algorithmic}[1]
\FOR {$\ell$ = $H-1$ downto $1$}  
\FORALL{nodes $i$ at level $\ell$}  
   \STATE $\x \gets PutInUpState (i, \x)$
\ENDFOR
\ENDFOR
\STATE \RETURN $\x$ \hfil \% \textit{UpSparse} Solution
\end{algorithmic}
\vspace{2mm}
\end{minipage} \\
\begin{minipage}{0.4\textwidth}
\vspace{2mm}
\textbf{Function} {$PutInUpState(i, \x)$}  \   \% Put  $i$ in upstate
\begin{algorithmic}[1]
\STATE $\delta \gets \min_{j \in c(i)} \{x_j\}$ 
\STATE $x_i \gets x_i + \delta$ 
\FORALL {nodes $j \in c(i)$}       
      \STATE $x_j \gets x_j - \delta$
\ENDFOR    
\STATE \RETURN $\x$  \hfil \% Link $i$ \& children updated
\end{algorithmic}
\vspace{2mm}
\end{minipage} \\
\hline
\end{tabular}
\caption{Algorithm {\bf UpSparse} for noiseless observation}  
\vspace{-2mm}
\label{algo:upsparse}
\end{figure}

We call the state of the loss vector tree after the application of the algorithm the \textit{UpSparse solution}.
An example of initial solution is the receiver solution $\x_I=0, \x_R=\y$.
Reading figure~\ref{fig:exampleKonetree} from right to left provides an example of the algorithm in action. 
We now give and prove its main properties.

\vspace{2mm}
\noindent\textbf{UpSparse Properties:}\\
 Let $\x^*$ be the output of UpSparse with input $\x$, then\\
\noindent 1.  Each local complex in $\x^*$  is in upstate (by construction). 

\smallskip\noindent 2. \emph{UpSparse Uniqueness}:  For fixed $\y$ and any feasible input $\x$, UpSparse outputs the same $\x^*(\y)$, defined 
$\forall i \in V \backslash \{\rr\}$ by ($\y_{R(i)}$ denotes the observations in subtree rooted at node $i$):
%
\begin{align}
        x^*_i = \gamma_i  \mbox{ if $f(i)=\rr$, \ else \ }&  x^*_i = \gamma_i - \gamma_{f(i)}, \\
~~~\mbox{where } \gamma_i = \min~\{~\y_{R(i)}~\} .\quad \nonumber
\label{unique}
\end{align}

\noindent\emph{Proof}:
For each $i \in V \backslash \{\rr\}$, let $\gamma_i$ be assigned the loss according to $\x$ over the path from $\rr$  to $i$. 
When the algorithm moves a complex centered at node $i$ to its upstate, by definition it subtracts $\delta = \min_{k \in d(i)} \{ x_k \}$ from each child link $k\in d(i)$ and adds it to the loss $x_i$ of link $i$. 
As a consequence $\gamma_k$  is unchanged $\forall k \in d(i)$ and $\gamma_i \gets \gamma_i + \delta = \gamma_i  + \min_{k \in d(i)} \{ \gamma_k - \gamma_i \} = \min_{k \in d(i)} \{ \gamma_k \}$ since $x_{k}=\gamma_{k}-\gamma_{i}$.
Now begin at level $H$ at the bottom of the tree where $\gamma_{r}=y_{r}$, $\forall r\in R$.
The algorithm applies the move to upstate to nodes level by level from $H-1$ upward (the order within level does not matter). 
From the property stated above,  the $\gamma_{i}$ at a given level are overwritten once only, when the algorithm reaches that level. It follows that on completion  $\gamma_i = \min_{k \in d(i)} \{ \gamma_k \}$ $\forall i \in V \backslash \{0\}$.
Since $\gamma_r = y_r$ for receiver nodes, and since the minimum of the minima of subsets is equal to the minimum over their union, this implies that $\gamma_i = \min\{ \y_{R(i)} \}$ $\forall i \in V \backslash \{0\}$. 
It immediately follows that $x^*_i = \gamma_i - \gamma_{f(i)}$ unless $f(i)=\rr$, in which case $x^*_i = \gamma_i$ 
since $R(\rr)=R(d(\rr))$.

\smallskip
\noindent 3. \emph{UpSparse solution has minimal $\ell_0$ and $\ell_1$ norms}:  For any solution $\x$,  $\norm{\x^*}{0} \le \norm{\x}{0}$  and  $\norm{\x^*}{1} \le \norm{\x}{1}$.\\
\noindent \emph{Proof}:  UpSparse moves (complexes centered on) internal nodes to upstate one by one from the bottom up, visiting each node only once.   
Let $\x^h$ denote the intermediate solution after $h$ nodes have been changed to upstate, $0 \le h \le |I|$, where $\x^0 = \x$ is the initial (arbitrary) feasible solution.  Let node $i$ be the node moved to upstate in step $h$. 
Changing $i$ to upstate affects only loss values of links incident on $i$. 
The number of lossy links incident on a node cannot increase when it is converted to upstate and so $\norm{\x^h}{0} \le \norm{\x^{h-1}}{0}$. 
For $\ell_1$, we observe that $\norm{\x^{h}}{1} = \norm{\x^{h-1}}{1} - (|d(i)|-1)\delta\le \norm{\x^{h-1}}{1}$ since  $\delta \ge 0$.
Since from the uniqueness property the algorithm terminates with $\x^{|I|} = \x^*$, it follows that the UpSparse solution obeys $\norm{\x^*}{0} \le \norm{\x}{0}$  and  $\norm{\x^*}{1} \le \norm{\x}{1}$. Since $\x$ is arbitrary, the result follows.

\smallskip
\noindent 4.   $\norm{\x^*}{0} \le m = |R|$.\\  
\noindent \emph{Proof}:  
For any $\y$, there always exists the receiver solution $\x_I=0, \x_R=\y$ with sparsity $|R|$. Thus $\norm{\x^*}{0} \le \norm{\x}{0} = |R|$.

\smallskip
\noindent 5. \emph{$\min \ell_1$uniqueness}:  UpSparse solution $\x^*$ is the unique solution with minimal $\ell_1$ norm.\\
\noindent\emph{Proof}:  
From 1 and 2, $\x^{*}$ is the only solution with all nodes in upstate. Therefore any other solution $\x$ must have at least one node $i$ which is not in upstate. Converting $i$ to upstate decreases $\norm{\x}{1}$ by $(|d(i)|-1)\delta>0$ since $i$ was previously not in upstate, implying that 
$\norm{\x}{1} > \norm{\x^*}{1}$. The result follows.

\smallskip
\noindent 6. \emph{$\min \ell_0$ uniqueness}: If each local complex in $\x^*$ is uniquely sparse,  then $\x^*$ is the unique sparsest solution, otherwise not.\\
\noindent \emph{Proof}:  Let $\x\ne\x^*$ be such that $\norm{\x}{0} = \norm{\x^*}{0}$. From 1 and 2, $\x^{*}$ is the only solution with all nodes in upstate, so $\x$ must have at least one node $i$ in downstate. Now $\x^* = UpSparse(\x)$, implying that at least one node in $\x^*$ can be changed to downstate to reach $\x$ at fixed sparsity, a contradiction, as all nodes of x* are uniquely sparse. \\
\emph{Corollary}: When the number of lossy links at  internal node $i$ with degree $g_i$ is at most $g_i-2$ $\forall i$, then the sparsest solution is unique. This degree condition ensures that no internal node can be moved from upstate without increasing sparsity, and hence $\x^* = UpSparse(\x)=\x$.

\noindent 7. \emph{Minimal $\ell_1$/UpSparse solution = true solution?}: If there exists at least one lossless child link at every internal node, UpSparse solution $\x^*$ is same as the true solution $\bar{\x}$.\\
\emph{Proof}:
From property 2, among all possible initial solutions, only UpSparse solution will pass through the UpSparse algorithm unchanged i.e. $\x^* = UpSparse(\x^*)$. Therefore the true solution $\bar{\x} = \x^*$ iff $\bar{\x} = UpSparse(\bar{\x})$. 
This occurs iff the true solution already has all its internal nodes in upstate. 
If there is at least one lossless child link at every internal node, no loss can be pulled up, thus all nodes are in upstate,  so $\x^* = \bar{\x}$. 

\subsection{Sufficiency condition for non-uniqueness in networks} 
\label{ssec:suffnet}

Consider a measurement matrix $A$ that defines a graph using paths from an underlying general network. Each column vector in $A$ corresponds to a link in the graph. For any branch or internal node, let $g^{in}$ (resp.~$g^{out}$) denote its in-degree (resp.~out-degree), that is the number of links covered by its incoming (resp.~outgoing) paths.  Then: 

{\bf Lemma 1}
\emph{
Let $K \ge \max \{g^{in}, g^{out}\}$  for some branch  node. 
Then there exists a $K$-sparse non-negative vector $\x$ that is not the unique sparsest solution to $\y(\x) = A\x$.}
\smallskip\\
\emph{Proof}:
We will prove by constructing two non-negative vectors $\bu$ and $\bv$ such that $A(\bu - \bv) = \zero$ with $\bu$ of sparsity $K$ and $\bv$ of sparsity at most $K$. This will imply that $\bu$ is not the unique sparsest solution of $\y(\bu) = A\bu$. Let $A = [\phi_1,\ldots,\phi_n]$ where $\phi_i$ is the $i$th column vector of $A$. Each column vector of $A$ corresponds to a link in the graph. Let $A^{in}$ (resp. $A^{out}$)  denote the set of all column vectors corresponding to links covered by incoming (resp. outgoing) paths. Now since each path that enters the branch point also exits it, the set of all paths covered by incoming links is exactly same as the set of all paths covered by outgoing links. Therefore 
$\sum_{\phi_i \in A^{in}}  \phi_i - \sum_{\phi_j \in A^{out}} \phi_j = 0$. In other words, $A^{in} \cup A^{out}$ is a dependent set of vectors and there exists a vector $\bh$ in the null space of $A$ such that $h_i = +1 ~\forall \phi_i \in A^{in}$, $h_i = -1 ~\forall \phi_i \in A^{out}$, and $h_i = 0 ~\forall \phi_i \not\in A^{in} \cup A^{out}$ so that $A\bh = \zero$. 

There are two ways to proceed from here. We could either directly apply theorem 1 of \cite{infocom} or provide a method to construct $\bu$ and $\bv$. We follow the later approach for ease of exposition. Without loss of generality let $g^{in} > g^{out}$ and $K \ge g^{in}$. We set $\bu, \bv$ as follows. $\forall \phi_i \in A^{in}$, $u_i = w > 0, v_i = 0$ so that $u_i - v_i = w$. We partition $A^{out}$ into two arbitrary subsets $A^{out}_1$ and $A^{out}_2$ of sizes $K-g^{in}$ and $g^{out} - (K-g^{in})$.  $\forall \phi_i \in A^{out}_1$, $u_i = w, v_i = 2w$ so that $u_i - v_i = -w$. $\forall \phi_i \in A^{out}_2$, $u_i = 0, v_i = w$ so that $u_i - v_i = -w$. $\forall \phi_i \not\in A^{in} \cup A^{out}$, $u_i = v_i = 0$. Therefore $\bu - \bv = w \bh$ or $A(\bu - \bv) = \zero$. By construction, $\norm{\bu}{0} = K$ and $\norm{\bv}{0} \le K$. 

Lemma~1 gives a worst case bound relating the degree of networks and their amenability to sparse recovery. If networks have small degree, then sparsity minimization can fail even for small $K$ if lossy links are concentrated at a branch point. 
For a binary tree, the bound for non-uniqueness given by lemma~1 is $K \ge 2$ since all branch nodes have $g^{in}=1$ and $g^{out}=2$. 
However property 6 shows that sparsity minimization will still succeed for a binary tree for higher $K$ provided the number of lossy links at each branch point is $<2$.


\section{The Noisy Problem}
\label{sec:noise}

The model defined in section~\ref{sec:model}  is based on knowing the 
mean loss observed at each receiver exactly.
This assumption can fail in a number of ways, the most important of which is 
that, in practice, loss is estimated based on a finite number of observations only, 
resulting in receivers seeing only an estimate $\hat{\y}$ of the true path loss observation vector $\y$.
Depending on how the measurements are made, 
the degree of dependence between the elements of $\hat{\y}$ will vary. 
In order to be as general as possible, we take the following non-parametric approach to the problem of coping with such `noise'.

To each receiver $j$ we will associate an interval 
$[y^\ell_j,y^u_j]$ 
within which the true value $y_j$ is deemed to lie.  Intuitively these intervals correspond to confidence intervals about the estimate $\hat{y}_j$ of $y_j$ for each $j\in R$.  We do not however ask how these confidence intervals are established, for example they could have arisen simply through an observation error at the receivers, rather than through too few probes. 
Instead, we take the conservative view that any 
$\y\in[\y^\ell,\y^u]$ (i.e.~$\forall j, y_j\in[y^\ell_j,y^u_j]$), is an equally possible observation vector. 
Let $\Xx$ denote the space of all possible solutions vectors $\x$ consistent with any $\y\in[\y^\ell,\y^u]$. 
$\Xx$ is now much wider, equal to the union of the null spaces arising from each possible $\y$.  We seek to understand how $\ell_0$ and $\ell_1$ regularisation operates in such a context. 
We regularise as follows: out of solutions with minimal $\ell_0$, we select the one with the smallest $\ell_1$.

\begin{figure*}[ht!]
\center
\includegraphics[width=\oneup]{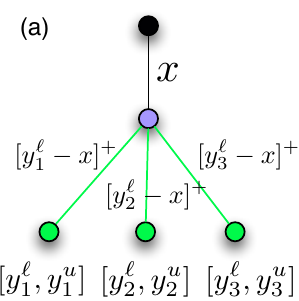}
\includegraphics[width=41mm]{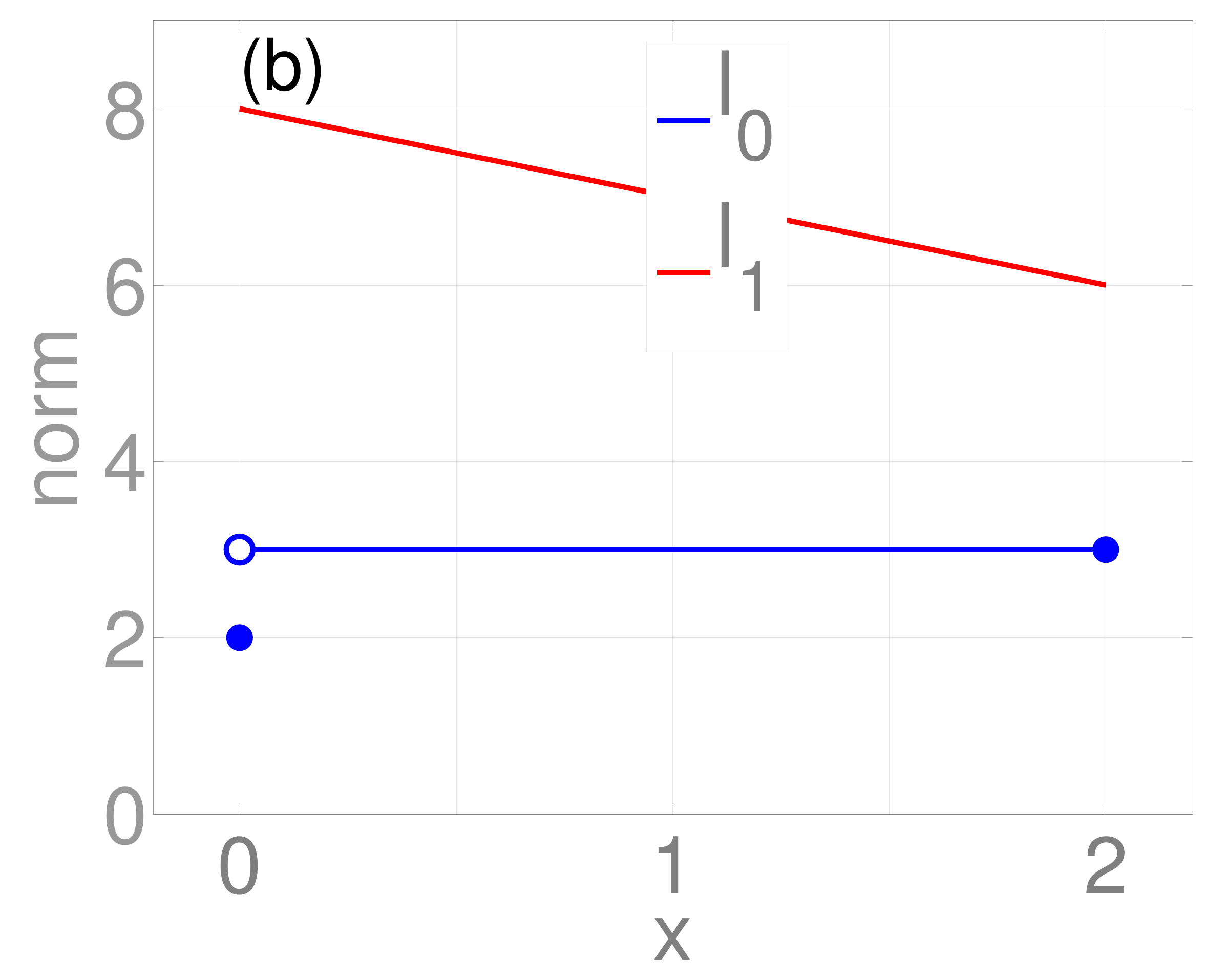}
\includegraphics[width=41mm]{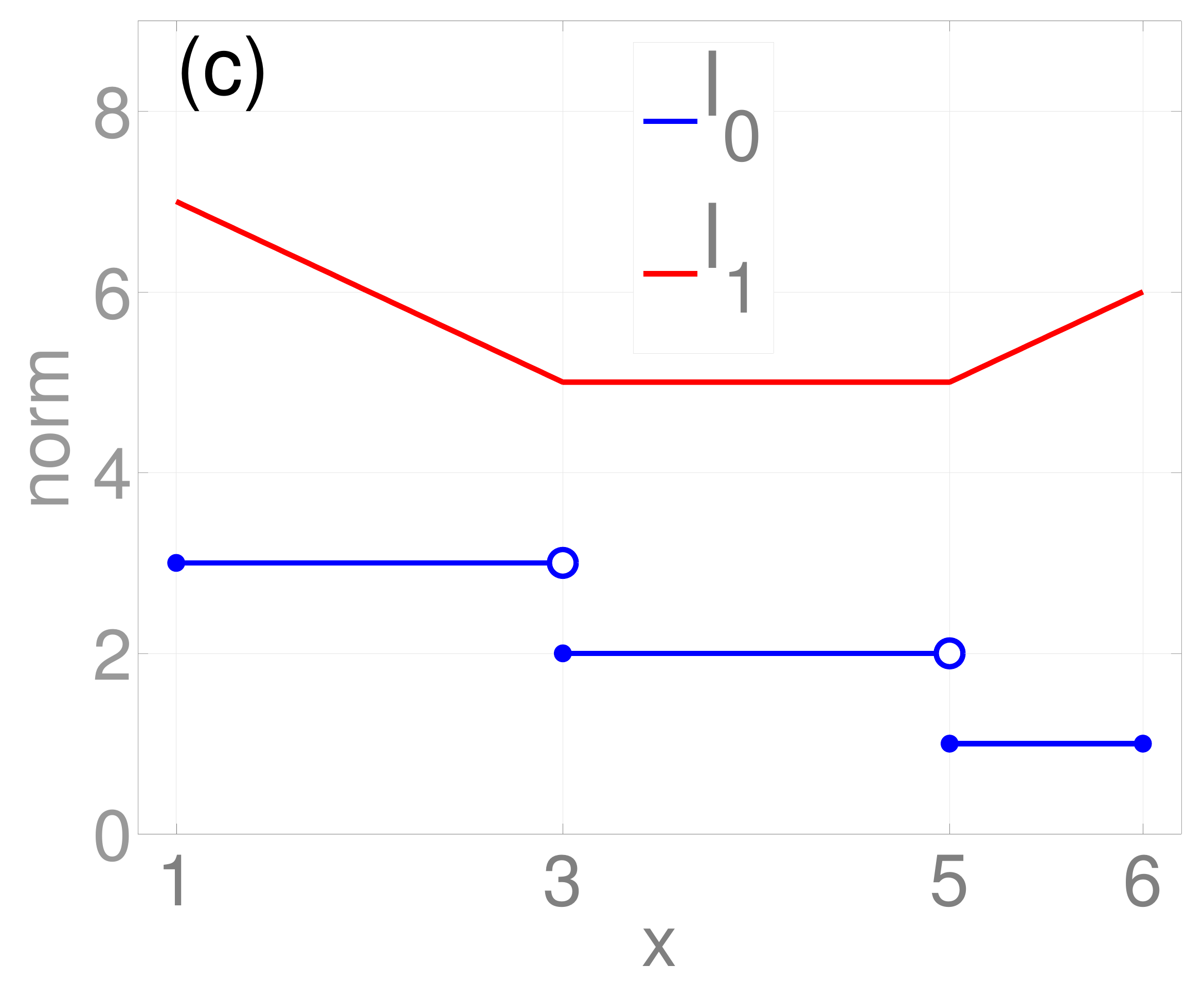}
\includegraphics[width=41mm]{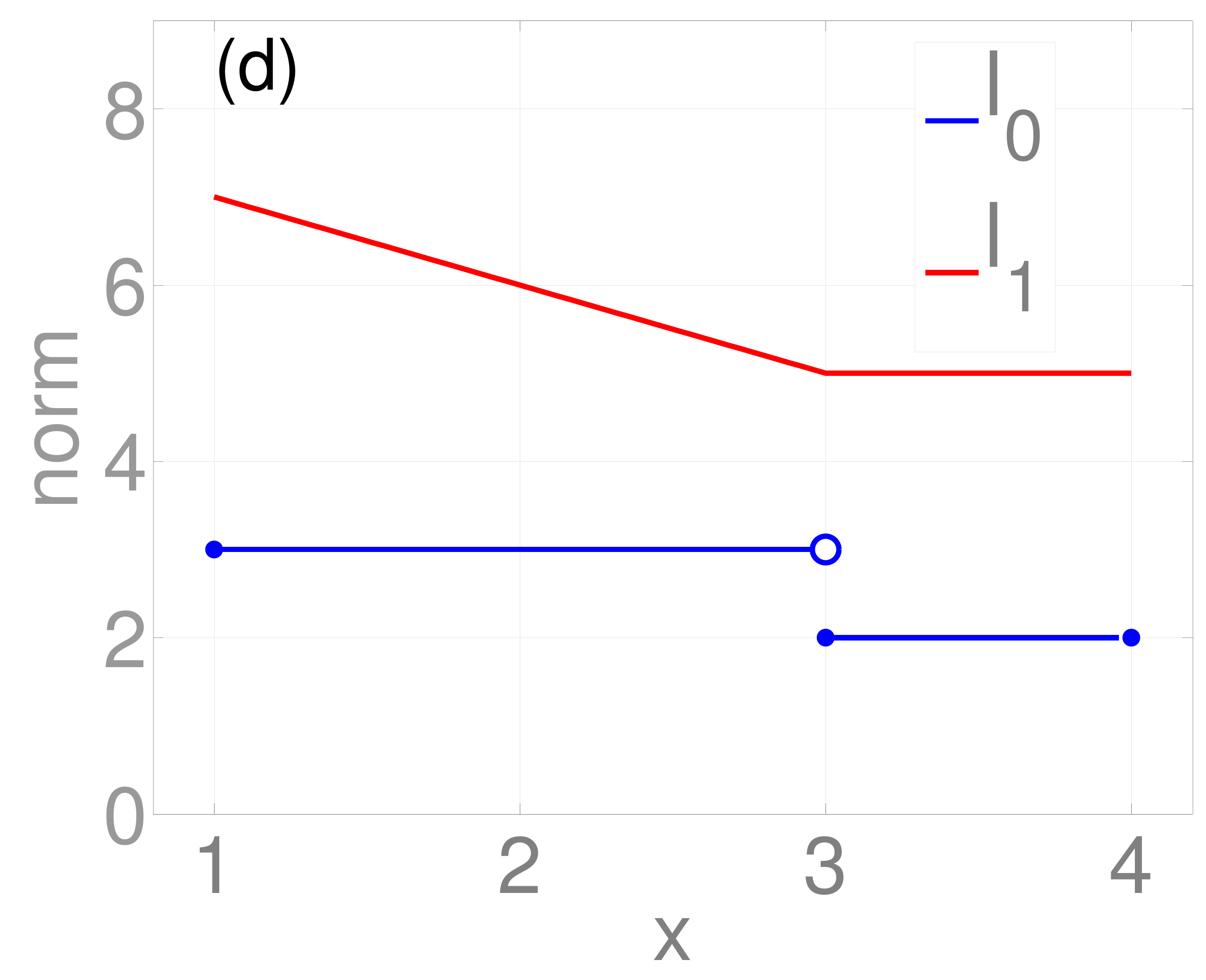}
\vspace{-0.3cm}
\caption{Noisy problem: Receiver observation $\y \in [\y^\ell, \y^u]$. Ambiguity at a local complex, which could be centered on any internal node, is illustrated using a $3$-receiver tree (left). The $\ell_0$ and $\ell_1$ norms of solutions $\x' = [\ [y^\ell_1-x]^+, [y^\ell_2-x]^+,  [y^\ell_3-x]^+, x\ ]$ are shown parametrised by $x$ for different examples: 
(b) $\y^\ell = [0, 3, 5],~ \y^u=[2, \infty, \infty]$: Minimal $\ell_0$ solution ($x=0$) is different from minimal $\ell_1$ solution ($x=2$), $y^L=0, y^U=2$.  
(c) $\y^\ell=[1,3,5],~ \y^u=[6, \infty, \infty]$: A unique solution ($x=5$) has both minimal $\ell_0$ and $\ell_1$ norms, $y^L=5, y^U=6$. 
(d) $\y^\ell=[1, 3, 5],~ \y^u = [4, 4, 6]$: Multiple solutions ($x \in [3,4]$) have both minimal $\ell_0$ and $\ell_1$ norms, $y^L=3, y^U=4$. }
\label{fig:noisy-ex}
\vspace{-0.5cm}
\end{figure*}

\subsection{Local Regularisation}
\label{ssec:localnoisy}

In this section we re-examine a local complex in the noisy context.
Whereas previously the general solution $\x' = [y_1-x, y_2-x, \ldots, y_m-x,  x]^T, \  x\in[0,\ymin]$, 
was a function of a single variable $x$ (see figure~\ref{fig:localtree}(a)), here the solution is a function of both $x$ and each element $y_j$ of $\y$.   
We first show how the problem can nonetheless be reduced to one parameterised by $x$ alone, by using a generalised notion of upstate.  
Let $\ylmin = \min_i \{ y^\ell_i \}$ and $\yumin = \min_i \{ y^u_i \}$. Similarly $\ylmax$, $\yumax$.

First, since the upstate solution  for any $\y$ fixed has minimal $\ell_0$ and $\ell_1$ norms, to find the minima 
over $\Xx$ it suffices to search over all upstate solutions.
In general the internal link takes values in the range $x\in[0,\yumin]$. 
For upstate solutions, this restricts to $x\in[\ylmin,\yumin]$.  
It is not hard to see that for fixed $x$ the path loss vector that yields the upstate solution with the lowest possible $\ell_0$ and $\ell_1$ norms is therefore
\be 
   \y'(x) = [ \max(x,y^\ell_1), \max(x,y^\ell_2), \ldots, \max(x,y^\ell_m)]^T.
\ee
This is because $y'_j=\max(x,y^\ell_j)$ is the smallest possible value in $[y^\ell_j,y^u_j]$ consistent with $x$ and the complex being in upstate.
The corresponding solution family, within which the minimal  $\ell_0$ and $\ell_1$ norms over $\Xx$  must lie, is (see also figure \ref{fig:noisy-ex}(a))
\be
 \x' = [\,[y^\ell_1-x]^+,  \ldots, [y^\ell_m-x]^+,  x ]^T,  ~~x\in[\ylmin,\yumin],
\label{eq:local-noisy}   
\ee
where $[x]^+=\max(0,x)$. To understand how this \textit{generalised local} or \textit{Glocal} solution compares to the noiseless case, let
$k(x)\in[0,m-1]$ be the number of $y^\ell_j$ strictly larger than $x$, which is also the number of lossy child links.   For these lossy links, increasing $x$ corresponds to moving loss from $x'_j>0$ up to the top link while holding $y'_j=y^\ell_j$ constant, just as in the noiseless case. For the other (lossless) child links where $y^\ell_j\le x$ we see something new: increasing $x$ implies that the $y'_j=x$ are themselves increased, with `new' loss being transferred immediately into the top link so that $x'_j$ remains zero. 

It remains to determine the optimal value of $x$ within the Glocal family for each norm, and to address uniqueness issues.  The key concept is that, as $x$ increases, it overtakes more and more of the $y^\ell_j$, so that the number of lossy links drops (lower $\ell_0$) and more loss is concentrated in the root link (lower $\ell_1$). \\\\
\noindent\textbf{Minimum sparsity and its uniqueness in a local complex}\\
\textit{Let $\yL=\max\{\ylj:\ylj\le \yumin\}$. The minimum $\ell_0$ in $\Xx$  is achieved by $\x^*_0$, the 
Glocal solution with $x=\yL$. $||\x^*_0||_0$ is either $k(\yL)+1$ or $k(0)$. \\
Case 1 ($\ylmin>0$): 
 \ the minimum is shared by all $x\in[\yL,\yumin]$, is unique if $y^L=\yumin$.  $||\x^*_0||_0=k(\yL)+1$.\\
Case 2 ($\ylmin=0$ and $\yL=0$):  
 \ the minimum solution is unique. $||\x^*_0||_0=k(0)$.\\
Case 3 ($\ylmin=0$ and $\yL$ is the 2nd smallest $y^l_j$): 
 \ the minimum solution is shared by $x\in\{0,[\yL,\yumin]\}$.  $||\x^*_0||_0=k(0)=k(\yL)+1$.\\
Case 4 ($\ylmin=0$ and $\yL$ is larger than the 2nd smallest $y^l_j$): 
 \ the solution is as in Case 1.
}
\smallskip\\
\noindent \emph{Proof}:
The number $k(x)$ of lossy child links is a monotonically decreasing function over $x\in[\ylmin,\yumin]$, and so its minimum is achieved at $x=\yumin$. It first reached this value the last time $x$ crossed a $\ylj$, hence it is shared by all $x\in[y^L,\yumin]$.\\ 
Case 1:  Since $x\ge\ylmin>0$, the sparsity is $k+1$ which attains its minimum where $k$ does.\\
Case 2:  Since $\ylmin=0$, $\ell_0$ jumps by 1 at the origin as $x$ moves above $0$. Since $\yL=0$, $k$ is constant over all $x$, and so the minimum is found at the origin, and is just $k$ since $x=0$.\\
Case 3:  As for Case 2, $\ell_0$ jumps by 1 at the origin. Since $\yL>0$ is the second largest, $k(\yL)=k(0)-1$ and so the sparsity at $\yL$ is $k(0)-1 +1=k(0)$, the same as at the origin.\\
Case 4:  From Case 3, if $\yL$ is the third smallest or greater, then the sparsity at the origin is no longer minimal and the solution is as in Case 1.

There is a remaining kind of non-uniqueness for $\ell_0$, namely that of the `downstate' type discussed under the noiseless case. However, since such cases always have larger $\ell_1$ norm, we will not consider them here.\\

\noindent\textbf{Minimum $\ell_1$ norm and its uniqueness  in a local complex}\\
\textit{Let $\yU=\min\{\ylmax, \yumin\}$. The minimum $\ell_1$ in $\Xx$  is achieved by $\x^*_1$, the 
Glocal solution with $x=\yU$ and $||\x^*_1||_1= x + \sum_j [\ylj - x]^+$.
The minimum is unique if $\yU < \bar y$, the 2nd largest $\ylj$, otherwise it is 
shared by all $x\in[\bar y,\yU]$.
}
\smallskip\\
\noindent \emph{Proof}:
It is easy to check that $||\x'(x)||_1$ is a piecewise linear function with slope $1-k(x)$. 
Since $k(x)=0$ for the first time at $x=\ylmax$, the minimum is found at the largest possible $x$ not exceeding this, i.e.~$\yU$.
If $\yU<\bar y$ then $k\ge2$, the slope is negative and so the minimum is achieved at $\yU$ only. 
Otherwise $k=1$, the slope is flat, and so the minimum is shared over $x\in[\bar y,\yU]$.

Figure \ref{fig:noisy-ex} shows three examples which illustrate the behavior of $\ell_0$ and $\ell_1$ norms in a local loss complex. Each example has different observation intervals $[\y^\ell, \y^u]$ and since solutions can be parametrised by $x$, the norms are plotted as a function of $x$. In~\ref{fig:noisy-ex}(b), $\y^\ell = [0, 3, 5],~ \y^u=[2, \infty, \infty]$. 
At $x=0=y^L$, the solution $\x' = [\ [y^\ell_1-x]^+, [y^\ell_2-x]^+,  [y^\ell_3-x]^+, x\ ] = [0, 3, 5, 0]$ and $\norm{\x'}{0}=2$, $\norm{\x'}{1}=8$. 
At $x=2=y^U$, $\x' = [0, 1, 3, 2]$ and $\norm{\x'}{0}=3$,  $\norm{\x'}{1}=6$. 
Hence the minimal norm solutions are $\x^*_0 = [0, 3, 5, 0]$ and $\x^*_1 = [0, 1, 3, 2]$. For \ref{fig:noisy-ex}(c), $\x^*_0 = \x^*_1 = [0, 0, 0, x=5]$. In \ref{fig:noisy-ex}(d), all solutions for $x \in [3,4]$ have minimal $\ell_0$ and $\ell_1$ norms. 

Special cases aside, the  \textit{generalised upstate} or \textit{Gupstate} solutions  
$\x^*_0$ and $\x^*_1$ are similar in many ways to their noiseless counterparts. 
In many cases $\x'$ with $x=y^U$ is optimal for each norm.  The addition of noise however has significantly expanded the number of cases where the optimal solutions are not unique.
As before, by preferring the solution where both norms find their minimum, these can be significantly reduced, though not eliminated as in the noiseless case.

\subsection{Global Regularisation}
\label{ssec:globalnoisy}

We now describe the {\bf$\gup$} algorithm for finding the minimal $\ell_0$ and $\ell_1$ solutions for arbitrary trees.
$\gup$ regards the subtree with path $(\rr, i)$ and child paths $\{(i, j): j \in R(i)\}$ essentially as a loss complex, defining
$\zli$, $\zui$, and $\zmi$ for a node $i$ as counterparts of $\yL$, $\yumin$, and $\ylmax$. 
For all $i \in V\setminus\{\rr\}$, define $\zui = \min \{ \yuj: j \in R(i)\}$, $\zli  = \max \{ \ylj: j \in R(i),  \ylj \le \zui\}$, and $\zmi = \max \{ \ylj: j \in R(i)\}$. Thus for leaves, $\z^L_R = \z^\ell_R = \y^{\ell}$, and $\z^u_R = \y^{u}$, and the full $\z^L$, $\z^u$, and $\z^\ell$ can be calculated recursively bottom up. For $\z^L$ the calculation is more complex. The $\ylj$ to the left of the corresponding $\yuj$ for each receiver set $R(c(i))$ of the children of $i$ must be retained to determine the new value. The complexity is $O(n^2)$.

In the algorithm,  $z_i \in [\zli, \zui]$ is assigned the loss of path $(\rr, i)$ in the same manner that $x \in [y^L, \yumin]$ holds the loss of the internal link in a loss complex. This ensures maximum number of lossless child paths under $i$. As $\gup$ moves top-down, it repeats the same for each node $i$, however leaving lossless paths set by parent nodes unchanged, effectively optimizing over the entire tree.

\begin{figure}[h!]
\vspace{0mm}
\begin{tabular}{l}
\hline
\begin{minipage}{0.4\textwidth}
\vspace{1mm}
\textbf{Function} {$\gup$}\\
$\x$: link solution,
$\z$: path solution
\begin{algorithmic}[1]
\STATE $z_{\rr} \gets 0$
\FOR {$p$ = $1$ to $H$} 
\FORALL{nodes $i$ at level $p$} 
\STATE $x_i \gets [\zli - z_{f(i)}]^+$
\STATE $z_i  \gets z_{f(i)} + x_i$
\ENDFOR
\ENDFOR
\STATE $\y = \z_R$
\STATE \RETURN ($\x, \y$)
\end{algorithmic}
\vspace{1mm}
\end{minipage}\\\hline
\end{tabular}
\caption{Algorithm {\bf $\gup$} for noisy observations}
\vspace{-0.1cm}
\end{figure}

\smallskip\noindent{\bf $\gup$ Properties}:

\smallskip\noindent 1. \emph{Minimal lossy links at each level in the tree}: $\forall$ nodes $i$, $x_i$ holds loss of link $i$ and $z_i \in [\zli, \zui]$ holds the loss of path $(0, i)$.  Given $\{ [\zlj, \zuj] :  j \in c(i) \}$, assigning: (\emph{a}) $z_i$ s.t. $\zli \le z_i \le \zui$ and (\emph{b}) $x_j = [\zlj - z_i]^+$, results in a minimal number of lossy child links of $i$, just as in a loss complex.

\smallskip\noindent 2. \emph{$\gup$ yields a minimal $\ell_0$ norm solution}.

\smallskip
\noindent \emph{Proof}:
We prove by induction on the height of tree. Let $\x(\T, [\y^\ell, \y^u])$ denote a feasible solution $\x$ for tree $\T$ given observation intervals $[\y^\ell, \y^u]$. Let $\x^*(\T, [\y^\ell, \y^u])$ denote the $\gup$ solution and $\z^*$ be the corresponding path solution.

\emph{Base case and induction hypothesis}: From \ref{ssec:localnoisy}, it is clear that $\gup$ is optimal for all trees of height $2$, i.e., given any observation interval $[\y^\ell, \y^u]$ and a tree $\T$ of height 2, $\x^*(\T, [\y^\ell, \y^u])$ has the lowest $\ell_0$ norm. Let us assume it is optimal for all trees of height $\le H$. 

\emph{Induction step}:  Consider a tree $\T$ of height $H+1$. Let $\T_H$ denote the tree consisting of all nodes and links of $\T$ from level $0..H$. So nodes of $\T$ at level $H$ are receivers of $\T_H$. 

Let $s = \norm{\x^*}{0}$. For each node $i$ at level $H$, $\gup$ assigns $z^*_{i} \in [\zli, \zui]$ and $\forall j \in c(i)$ $x^*_j = [\ylj - z^*_i]^+$. Let this result in a total of $s_1$ lossy links at level $H+1$. By property $2$, $s_1$ is minimal. 

Let $\z_H$ denote the part of $\z$ assigned to nodes at level $H$. Let $\x_H$ denote the part of $\x$ assigned to tree $\T_H$. By induction hypothesis, 
$\x_H^*(\T, [\z^L_H, \z^u_H])$ is optimal. Let $\norm{ \x_H^*}{0} = s - s_1$. 

Let $\x'(\T, [\y^\ell, \y^u])$ be an alternate solution that assigns $z'_i$ to node $i$ at level $H$. Let $\norm{\x'}{0} =  s'$ and let $s'_1$ be the number of lossy links on level $H+1$. We have two cases: 

\noindent(\emph{i}): If $z'_{i} \in [\zli, \zui]$ $\forall i$ at level h, then $s'_1 = s_1$. By induction hypothesis, 
$\norm{\x_H'(\T_H,  [\z^L_H, \z^u_H] )}{0} \ge s - s_1$.  So $\norm{\x'}{0} \ge s$.

\noindent(\emph{ii}):  $z'_{i} \not\in [\zli, \zui]$ for some node $i$ at level $H$. Since $z'_i > \zui$ makes $\x'$ infeasible, it must be that $z'_i < \zli$. For contradiction, let $s' < s$. Now $\x'$ can be modified by making (a) $z'_i = \zli$ and (b) $\forall j \in c(i)$, $x_j = [y^\ell_j - z'_i]^+$. This makes at least one child link of $i$ lossless and link $i$ could become lossy if it was lossless earlier. Since we gain at most one lossy link and loose at least one, $s'$ either decreases or remains unchanged. So still $s' < s$. Doing this for all nodes $i$ at level $H$ with $z'_i < \zli$ results in $s'_1 = s_1$. After this step, since $z'_i \in [\zli, \zui]~ \forall i$ at level $H$, $\x_H'$ is a solution for $\T_H$  w.r.t observation interval $[\z^L_H, \z^u_H]$. By induction hypothesis $\norm{\x_H'(\T_H,  [\z^L_h, \z^u_h] )}{0} \ge s - s_1$ i.e. $s' - s'_1 \ge s - s_1 \Rightarrow s' - s_1 \ge s - s_1 \Rightarrow s'  \ge s$ which is a contradiction.

\smallskip\noindent 3. \emph{$\gup$ will find the minimal $\ell_1$ solution if $\zli$ is replaced by $\min\{ \zmi, \zui \}$ in line $4$}. This follows simply from the fact that $\min\{ \zmi, \zui \}$ is the analogue of $\yU$ for a loss complex. 

\smallskip\noindent 4. \emph{$\gup$ will find the minimal $\ell_1$ solution among minimal $\ell_0$'s if the condition below is inserted between lines $4$ and $5$}. 
\begin{algorithmic}
\IF { ($x_i > 0$) and ($\zui <  \zmi$) } 
\STATE $x_i \gets \zui - z_{f(i)} $
\ENDIF
\end{algorithmic}
From ~\ref{ssec:localnoisy}, in a loss complex whenever $\yumin < \ylmax$, moving from $x = y^L$ to $x = \yumin$ reduces $\ell_1$ or keeps it minimal. This also keeps $\ell_0$ minimal provided $y^L > 0$. The condition above simply tests this and increases the loss of link $i$ provided minimal $\ell_0$ solution had not required it to be $0$. After this step, still $z_i \in [\zli, z^u_{\min}]$, therefore lossless child links of $i$ are unaffected. 

\smallskip\noindent 5. When $\y = \y^\ell = \y^u$, the $\gup$ solution is same as the UpSparse solution for $\y$. 

\smallskip\noindent 6. The function $\gup$  performs $O(n)$ \!steps. However setting $\z^L$, $\z^u$ and $\z^\ell$ takes $O(n^2)$ time. $\gup$ can be implemented using preorder and postorder traversals through the tree: postorder traversals to set $\z^L$, $\z^u$ and $\z^\ell$, and  preorder traversal to sets $\x$ and $\z$.

\section{Comparisons}
\label{sec:comp}

We now compare UpSparse to other optimisation methods from CS, of higher computational complexity, which could be used to find solutions with minimal $\ell_0$ or $\ell_1$ in trees. 
\begin{align}
\label{eq:l0-min}
\text{Noiseless:~} & \underset{x}{\min}~ ||\x||_0  :   A\x = \y,  x_i>0 \ \forall i\\
\label{eq:l1-min}
& \underset{x}{\min}~ ||\x||_1  :   A\x = \y,  x_i>0 \ \forall i\\
\label{eq:l1-l2}
\text{Noisy:~} & \underset{\x}{\min} ~\frac{1}{2}\norm{\y - A\x}{2}^2 + \lambda \norm{\x}{1} \\
\label{eq:l1-l1}
& \underset{\x}{\min} ~\norm{\y - A\x}{1}  + \lambda \norm{\x}{1} \\
\label{eq:l1-l1-prior}
& \underset{\x}{\min} ~\norm{\y - A\x}{1}  + \lambda \norm{\x - \pmb{\mu}}{1} 
\end{align}
In the noiseless case, the optimal sparsity optimisation problem is (\ref{eq:l0-min}) which is non-convex.  In general such problems are NP-hard.  
In contrast, UpSparse provides a low cost single pass algorithm, which achieves global optimality through only $O(n)$ local operations, the number of links in the tree. 
%
In the $\ell_1$ case, the optimisation problem is (\ref{eq:l1-min}) which is not only convex but a linear program.
Known as the basis pursuit formulation~\cite{bp}, it is used as a substitute for (\ref{eq:l0-min}).
Although its solution is straightforward using linear programming, UpSparse offers a low cost direct alternative which fully exploits the underlying tree topology.

The noisy problem has been tackled using the approaches (\ref{eq:l1-l2}-\ref{eq:l1-l1-prior}). Eq.(\ref{eq:l1-l2}) is the basis pursuit denoising formulation and the unconstrained Lasso \cite{lasso}  formulation. Eq.~(\ref{eq:l1-l1}) is used as an alternative to (\ref{eq:l1-l2}) as it is a linear program. It is used in~\cite{netquest}. 
Eq.~(\ref{eq:l1-l1-prior}) is proposed in \cite{netquest} and used in \cite{netscope} to choose a 
solution close to a prior $\pmb{\mu}$.
Compared to these approaches which require the introduction of a penalty term which is traded off with the $\ell_1$ norm (used to approach $\ell_0$) as well as a tradeoff parameter,  $\gup$ uses $\ell_0$ directly, supplemented by $\ell_1$ when needed. 
A relevant special feature here of the tree problem is that \textbf{any} noisy observation $\hat\y$ has a  feasible solution (for e.g. the receiver solution).
There is no need for regularisation in the sense of finding the closest feasible solution to $\hat\y$.

\section{Experiments}
\label{sec:exp}
\begin{figure}[h!]
\center
\includegraphics[width=53mm]{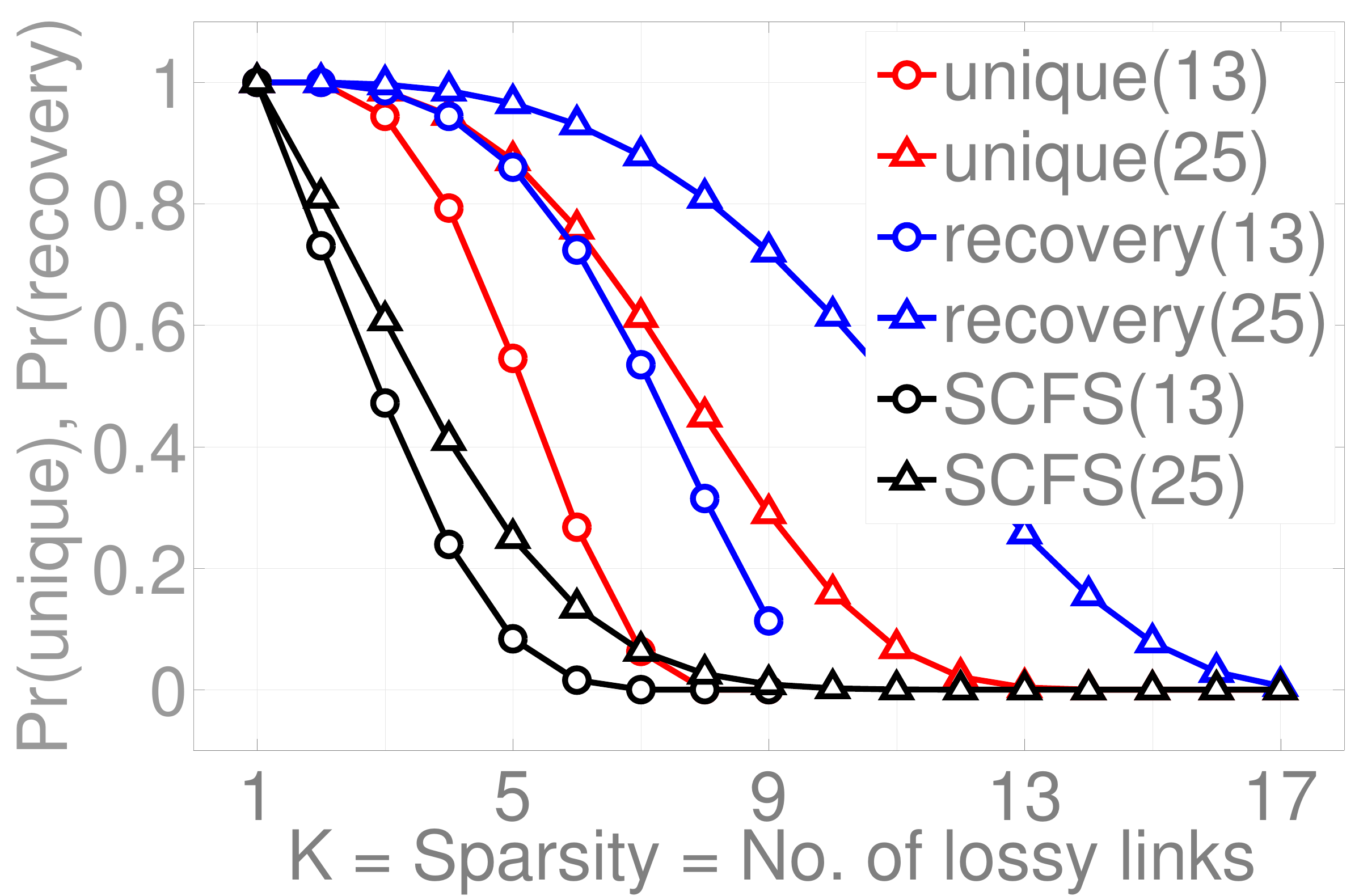}
\caption{Probability that minimal $\ell_0$ is unique and probability that minimal $\ell_1$ is the true underlying solution.}
\label{fig:ur}
\vspace{-0.3cm}
\end{figure}

\begin{figure*}[t!]
\center
\includegraphics[width=50mm]{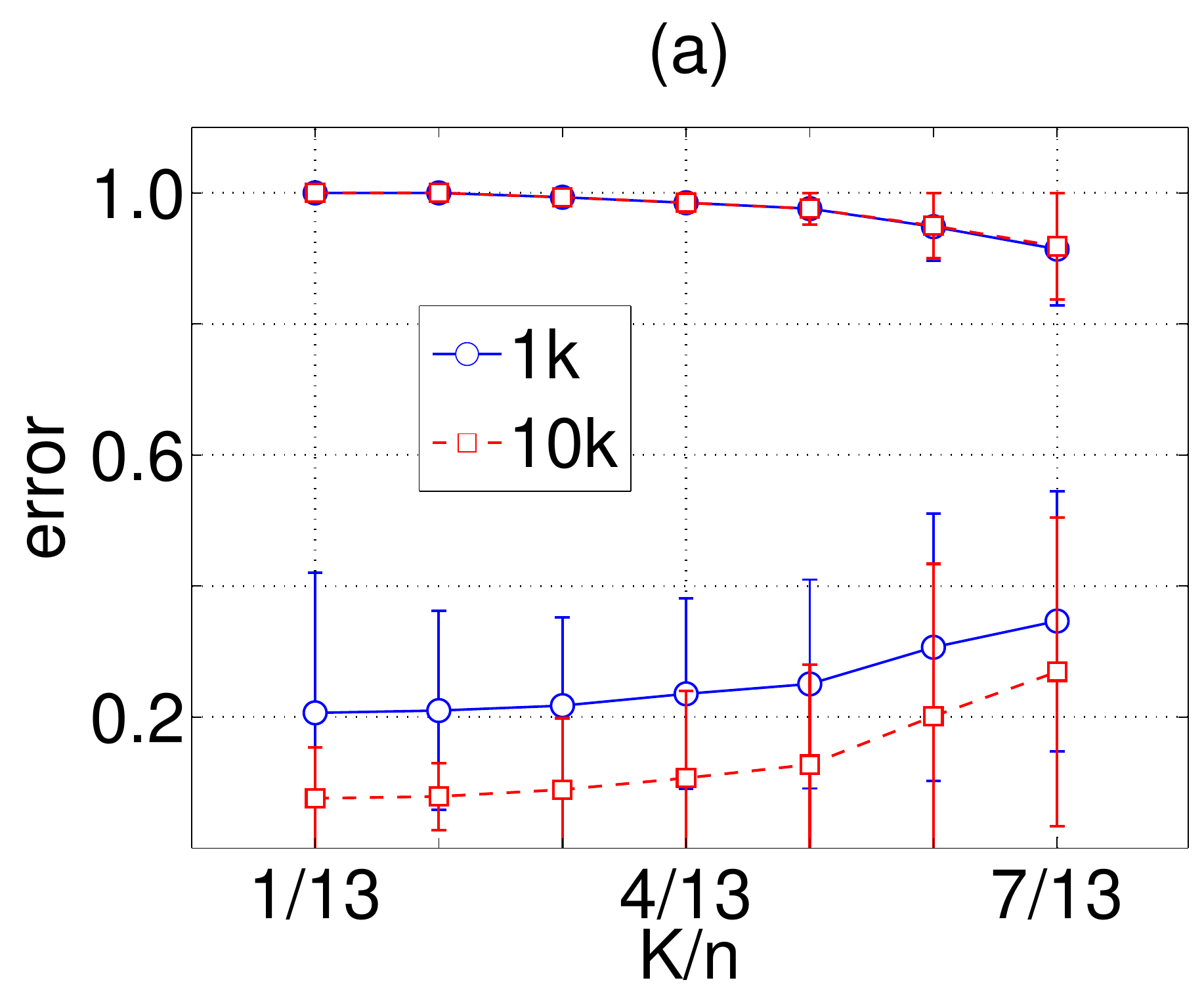}
\includegraphics[width=50mm]{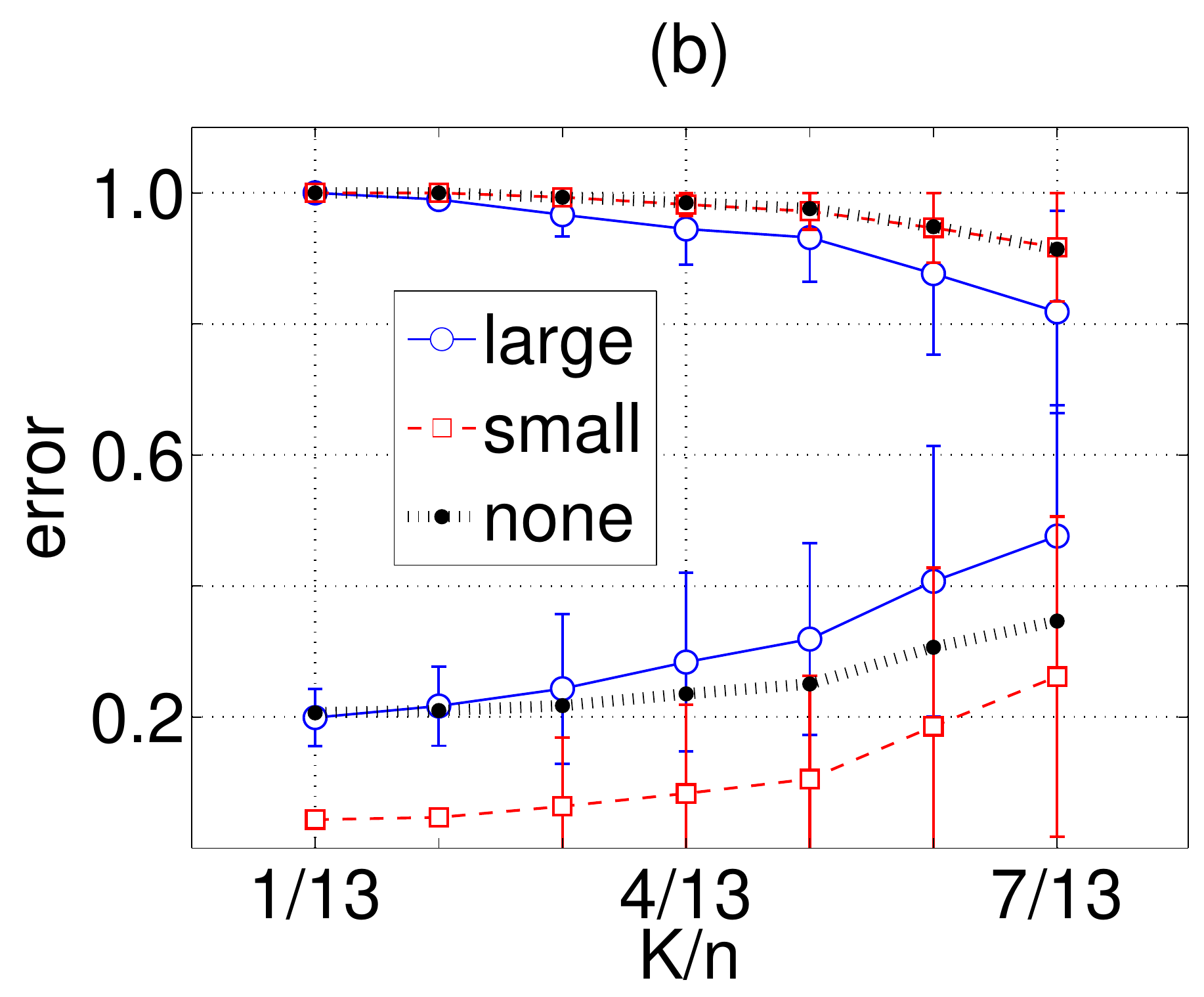}
\includegraphics[width=50mm]{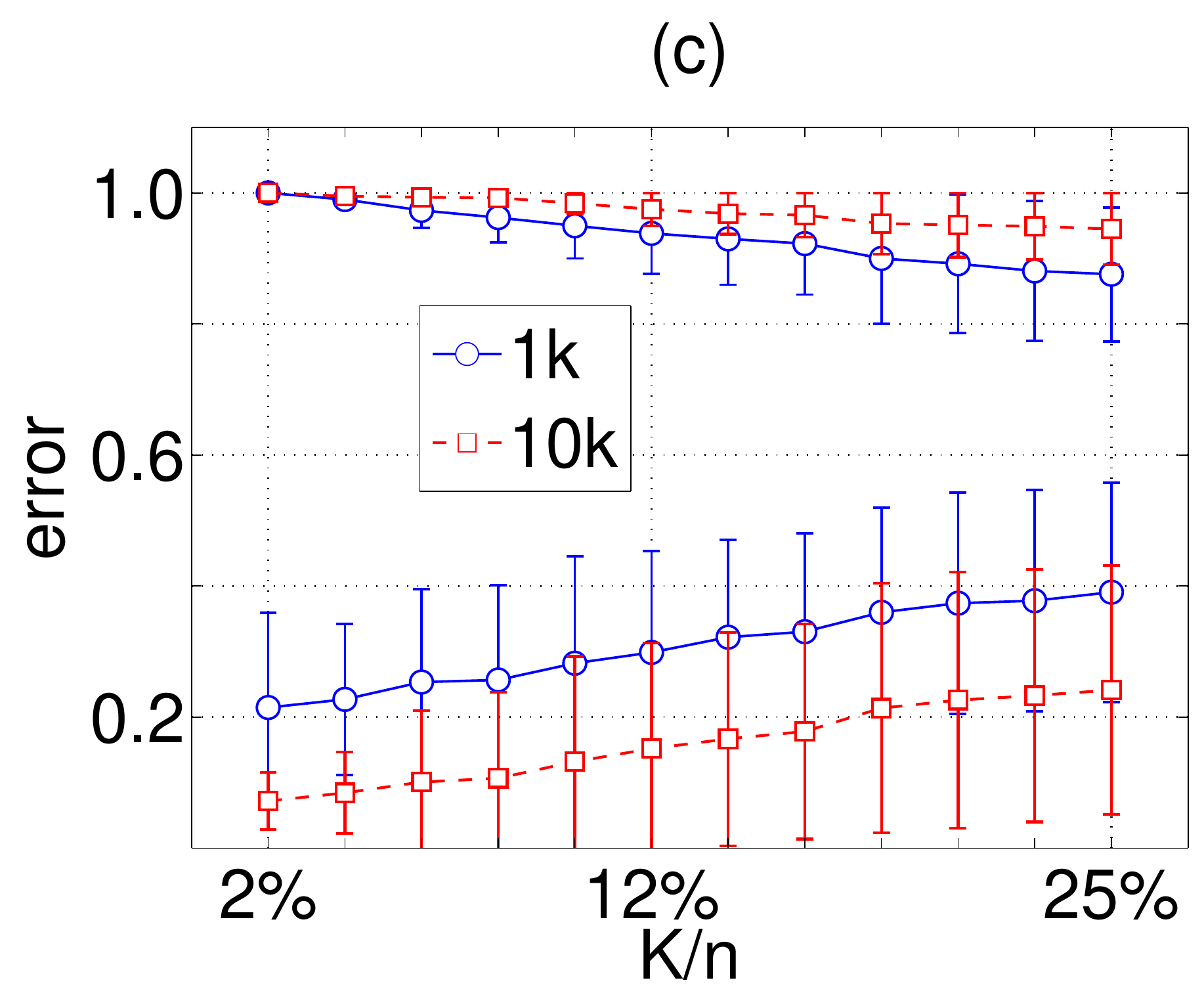}
\caption{
Effect of measurement noise for trees as a function of sparsity $K$ and the number of probes. Top curves show the success rate $e_0$ and bottom curves show relative $\ell_2$ error $e_2$ (a, b): Results for UpSparse and $\gup$ solutions respectively for a ternary tree with $n=13$ links, and (c): Results for realistic trees cut from AT\&T network.} 
\label{fig:exp}
\vspace{-4mm}
\end{figure*}

 This section summarizes our experimental results. 

\noindent\emph{Uniqueness of minimal $\ell_0$ and effectiveness of $\ell_1$ recovery in trees}: 
Figure \ref{fig:ur} 
plots different probabilities of interest as a function of sparsity $K$ by exhaustively looking at all ${n \choose K}$ links in trees. 
These are shown using two ternary trees with $n=13$ ($m=9$ leaves) and $n=25$ ($m=17$ leaves) links for $1 \le K \le m$. 
The red curves show the probability that the sparsest solution is unique. The blue curves show the probability that the minimal $\ell_1$ solution is the true underlying solution. 
%
%
%
%
Property $6$ of section \ref{ssec:global} gives the condition for uniqueness of sparsest solution and $7$ gives the condition that minimal $\ell_1$ = true solution. For $K \le 2$, minimal $\ell_0$ is guaranteed to be unique since for all internal nodes, the node degree $g=4$ (3 children, 1 parent) and $K \le g-2$ holds. For $K>2$, the probabilities gradually decay with increasing number of lossy links. For $K>2$, a vector of sparsity $K$ need not be unique in general (lemma 1). It is unique however if corollary of property $6$ is satisfied. We see that $\ell_1$ minimization can recover the true solution even when the sparsest solution is not unique. Since $\ell_1$ picks one of the sparsest solutions that is in upstate, when the true solution is the upstate solution, $\ell_1$ minimization recovers it. 
It is clear that UpSparse or any $\ell_1$ minimization algorithm effectively recovers the true loss hotspots in a tree, provided that only few of them exist.


We compare UpSparse to SCFS (smallest consistent failure set) algorithm~\cite{binary}, which recovers the sparsest link binary vector (each link either good/lossless or bad/lossy) given the path binary vector (each path either good or bad). 
The black curves show the probability that SCFS recovers the true link binary vector \emph{i.e.} all the locations of bad/lossy links. 
We see that SCFS has a lower success rate than UpSparse (blue curves) even though UpSparse recovers both the locations as well as loss rates of all lossy links. The binary approach yields higher ambiguity than the loss approach. For example, at a branch node, if one of the child links and the parent link are both bad/lossy, SCFS will report only the parent as bad. However UpSparse will report both links as lossy. 


\noindent\emph{Effect of measurement noise}: \ Next we conduct probing experiments where instead of the true path probabilities only an estimate is available, based on a fixed number of probes. We compute the minimal norm solutions and plot errors as a function of both sparsity and increasing number of probes. We assume a scenario where we have no prior knowledge of where the lossy links may lie. Thus we pick $K$ links uniformly at random. For each of these, loss is set uniformly at random from $1-10$\%. For the remaining links, loss is set to $0$. 
Using $\x$, 
we simulate the passage of probes on each path 
and derive the noisy observation $\hat{\y}$ and its confidence intervals $\hat{\y}^\ell$, and $\hat{\y}^u$. These are used by $\up$ and $\gup$ to yield $\hat\x$. Finally we recover the true loss estimates 
$\hat{\lossvec} = {\llog}^{-1}(\hat{\x})$. 

We compute two quantities: (\emph{i}) relative $\ell_2$ error $e_2 = \norm{\lossvec - \hat\lossvec}{2} / \norm{\lossvec}{2}$ and (\emph{ii}) Success rate $e_0$ which is the number of common lossy links between $\lossvec$ and $\hat\lossvec$ normalised by $\norm{\lossvec}{0}$. $e_0$ attempts to determine if sparsity can help identify the correct locations of lossy links. 

\noindent\emph{1) UpSparse Solutions}:\  Figure \ref{fig:exp}(a) shows benchmark results using a ternary tree of height $3$ with $n=13$ links. The top curves show the success rate $e_0$ and bottom ones show $e_2$ averaged over $100$ repetitions for $1$k and $10$k probes when $\up$ is given the noisy $\hat\y$ in each repetition. We see that even with $1$k probes, UpSparse picks the correct locations of lossy links with high probability. As probes increase, $\hat\y$ approaches $\y$ and $e_2$ decreases.

\noindent\emph{2) $\gup$ Solutions}: \ Figure \ref{fig:exp}(b) shows results for $\gup$ using $1$k probes for intervals of small and large sizes. In each repetition, $\gup$ is given intervals $\hat{\y}^\ell$, and $\hat{\y}^u$ that always contain the true $\y$. When interval sizes aren't too large, $\gup$ can find solutions which are even sparser than $\x$ resulting in higher error. As intervals get narrower, $\gup$ gives better results. The curve in the centre shows results when the actual noisy observation is used by $\up$. 

\noindent\emph{3) Large Realistic trees}: \ Finally~\ref{fig:exp}(c) shows results for real tree topologies cut out from the publicly available router level map of the AT\&T network obtained by Rocketfuel~\cite{rocket}, with about 48 links on average per tree. The figure shows results for solutions computed using $\gup$ for $1$k and $10$k probes when $\hat{\y}^\ell$, and $\hat{\y}^u$ are $t$-distributed intervals for $90$\% confidence, centered around $\hat\y$. We see that $e_2$ increases as expected since large trees will need more probes to get proper estimates of path probabilities. However the success rate $e_0$ remains high which implies that minimal norm solutions often identify the correct locations of lossy links. 

\section{Conclusion}

The sparsity principle has been mostly used as a black box in loss tomography without detailed characterization of conditions under which minimal norm solutions are unique or recover the true underlying solution. These conditions are important in practice as network operators wish to know when sparsity could be used to accurately localise hotspots using few monitoring points. 

In this work, we study the problem of loss hotspot localization in tree topologies (e.g. server-based measurements) using the principle of sparsity. 
We derive explicit solutions and fast algorithms for both min $\ell_0$ and  $\ell_1$ norms that give deep insight into the nature of sparsity in trees. 
We provide conditions under which minimal norm solutions are unique and when they recover the true underlying solution. 
We show that when lossy links are well separated, sparse solutions remain unique in many cases. 
We conduct experiments to measure the ability of the minimally sparse solution to approach the actual sparse solutions in practice. We see that minimal norm solutions can identify the locations of most lossy links, however as their number increases it becomes much harder to identify true loss rates. 
We also observe that minimally sparse link loss solutions can localize hotspots better than minimally sparse link binary solutions used in prior work. 
%
%

Future work will extend our work to graphs, study recovery conditions for the binary performance problem, and test our results with real measurements. 




\bibliographystyle{IEEEtran}
\bibliography{ref}

\end{document}